\documentclass[lettersize,journal, twoside]{IEEEtran}
\usepackage{amsmath,amsfonts}
\usepackage{array}
\usepackage[caption=false,font=footnotesize,labelfont=rm,textfont=rm]{subfig}
\usepackage{textcomp}
\usepackage{stfloats}
\usepackage{url}
\usepackage{verbatim}
\usepackage{graphicx}
\usepackage{cite}
\usepackage{url}

\usepackage{amsmath,amssymb,amsfonts}
\usepackage{algorithmic}
\usepackage[linesnumbered,ruled]{algorithm2e}
\usepackage{enumitem}
\usepackage{booktabs}
\usepackage[dvipsnames]{xcolor}
\PassOptionsToPackage{hyphens}{url}
\usepackage{tikz}
\usetikzlibrary{arrows.meta,calc}

\usepackage{hyperref}
\hypersetup{colorlinks, citecolor=blue, filecolor=pink, linkcolor=red, urlcolor=magenta }
\usepackage[inkscapelatex=false]{svg}
\graphicspath{{./figure}}
\usepackage{textcomp}
\usepackage{xcolor,colortbl}
\definecolor{mygray}{gray}{0.95}
\definecolor{magicmint}{rgb}{0.85, 0.96, 0.92}
\definecolor{deployblue}{HTML}{DCEBFA}
\definecolor{deployblueframe}{HTML}{6FA6D8}
\definecolor{deploygreen}{HTML}{E4F2D9}
\definecolor{deploygreenframe}{HTML}{7CB56D}
\definecolor{deployorange}{HTML}{FFF0C7}
\definecolor{deployorangeframe}{HTML}{E3AC36}
\definecolor{deploygray}{HTML}{F4F4F4}
\usepackage{pifont}
\usepackage{fontawesome5}
\usepackage{multirow}

\newcommand{\tabincell}[2]{\begin{tabular}{@{}#1@{}}#2\end{tabular}}

\def\BibTeX{{\rm B\kern-.05em{\sc i\kern-.025em b}\kern-.08em
    T\kern-.1667em\lower.7ex\hbox{E}\kern-.125emX}}

\usepackage{lipsum}
\usepackage[htt]{hyphenat}
\usepackage{listings}
\usepackage{wrapfig}

\usepackage{enumitem}
\graphicspath{{./figure}}
\DeclareRobustCommand{\tabincell}[2]{\begin{tabular}{@{}#1@{}}#2\end{tabular}}

\usepackage{xspace}
\makeatletter
\DeclareRobustCommand\onedot{\futurelet\@let@token\@onedot}
\def\@onedot{\ifx\@let@token.\else.\null\fi\xspace}
\usepackage{multirow}

\def\etal{\textit{et al}\onedot}
\makeatother

\usepackage[most]{tcolorbox}

\usepackage{pifont}
\newcommand{\revise}[1]{{\color{black}{#1}}}
\newcommand{\summary}[1]{
\begin{center}
\begin{tcolorbox}[colback=gray!15, colframe=black, boxsep=-0.15cm, middle=-0.15cm,breakable]
\textbf{\ding{45} Summary}
$\vartriangleright$
{#1}
$\vartriangleleft$
\end{tcolorbox}
\end{center}
}

\usepackage{graphicx}
\usepackage{listings}
\usepackage{multicol}

\begin{document}

\title{Software Engineering Practice in the Virtual Reality/Augmented Reality Software Life Cycle: A Systematic Survey}


\author{Hanyang Guo, Jingwen Wu, Zhengyang Zhu, Gengyang Xu, Hong-Ning Dai,~\IEEEmembership{Senior Member,~IEEE}, Xiapu Luo,~\IEEEmembership{Senior Member,~IEEE} and Zibin Zheng,~\IEEEmembership{Fellow,~IEEE}
\thanks{H. Guo is with Sun Yat-sen University, Zhuhai 519082, China and Hong Kong Baptist University, Hong Kong, 999077, China   (e-mail: guohy36@mail2.sysu.edu.cn).

J. Wu is with Hong Kong Baptist University, Hong Kong, 999077, China (e-mail: csjwwu@comp.hkbu.edu.hk). 

Z. Zhu is with Sun Yat-sen University, Zhuhai 519082, China and Pengcheng Laboratory, Shenzhen, 518055, China (e-mail: zhuzhy57@mail2.sysu.edu.cn

G. Xu is with The Hong Kong University of Science and Technology, Hong Kong, 999077, China (e-mail: gxuah@cse.ust.hk). 

H.-N. Dai is with Hong Kong Baptist University, Hong Kong, 999077, China (e-mail: hndai@ieee.org). 

X. Luo is with The Hong Kong Polytechnic University, Hong Kong, 999077, China (e-mail: csxluo@comp.polyu.edu.hk). 

Z. Zheng is with Sun Yat-sen University, Zhuhai 519082, China (e-mail: zhzibin@mail.sysu.edu.cn).
}
\thanks{Corresponding Author: Hong-Ning Dai}
}


\markboth{Preprint---Submitted to IEEE Transactions on Software Engineering, July 2026}%
{Guo \MakeLowercase{\textit{et al.}}: Software Engineering Practice in the Virtual Reality/Augmented Reality Software Life Cycle}


\maketitle

\begin{abstract}
Leveraging diverse virtual reality (VR) technologies, VR provides users with an immersive experience in a computer-generated 3D virtual environment. In contrast, augmented reality (AR) superimposes virtual objects (such as images, text, animations, etc.) onto users' real scenes. Both VR and AR have received increasing attention due to the maturity of computer graphics, computer vision, and artificial intelligence. VR and AR have gained rapid market growth in both hardware and software. In the early stages, VR/AR-related research mainly focused on hardware (e.g., head-mounted displays and sensor-equipped controllers). Recently, VR/AR software has become a hot topic due to the mature VR/AR ecosystem. However, there is no comprehensive work to summarize software engineering practices in VR/AR. 

To bridge this gap, this paper aims to provide a systematic survey summarizing the state-of-the-art research on VR/AR from the perspective of software engineering practices. Considering the general VR/AR software life cycle, we categorize the relevant research based on the VR/AR software life cycle, including requirement engineering and planning, software design, software development, software testing, software deployment, software maintenance and updates, and software evaluation and feedback. As for each stage above, we summarize relevant research on VR/AR for this stage, including the design of technical routes, the construction of datasets, and the design of evaluation metrics. We also enumerate existing empirical studies and make comparisons with conventional software (e.g., mobile/desktop application development). In summary, our article can help researchers gain a comprehensive understanding of the achievements of the existing software development techniques in the VR/AR field and also provide insights, which may facilitate the benign development of VR/AR software ecosystem in the future.
\end{abstract}


\section{Introduction}

\IEEEPARstart{A}{s} emerging technologies connecting the virtual world and the real world, virtual reality (VR) and augmented reality (AR)~\cite{ElMiedany2019} have received extensive attention recently. Through a computer-generated simulation of the environment using visual, auditory, and other multi-sensory elements, VR provides users with an immersive virtual experience when they are wearing VR headsets (e.g., Meta Quest, HTC Vive, Apple Vision Pro, etc.). In contrast, AR superimposes virtual objects (e.g., images, sounds, text, etc.) onto the real world. Through diverse AR devices (e.g., smartphones, tablets, AR glasses, and VR headsets), users can see and interact with virtual elements in their actual environment, thereby enhancing their perception and understanding of the real world. Typical AR applications include the Pokémon GO game and IKEA Place. In recent years, with the rising popularity of the metaverse concept and the maturity of generative artificial intelligence (AI) and large language models (LLMs), the VR and AR markets have experienced explosive growth. The global VR market size valued at USD 20.83 billion in 2025 is projected to grow from USD 26.71 billion in 2026 to USD 171.33 billion by 2034, exhibiting a CAGR of 26.20\% during the forecast period (2026-2034)~\cite{Fortune}. The AR market valued at USD 140.34 Billion in 2025 is poised to grow from USD 210.96 Billion in 2026 to USD 2,344.90 Billion by 2034, growing at a CAGR of 35.10\% during the forecast period (2026-2034)~\cite{Fortune2}. In summary, VR and AR have received a rapid growth in interest from both industry and academia.

\begin{figure*}[!t]

\centering
\subfloat[Year \& Field]{\includegraphics[width=2.8in]{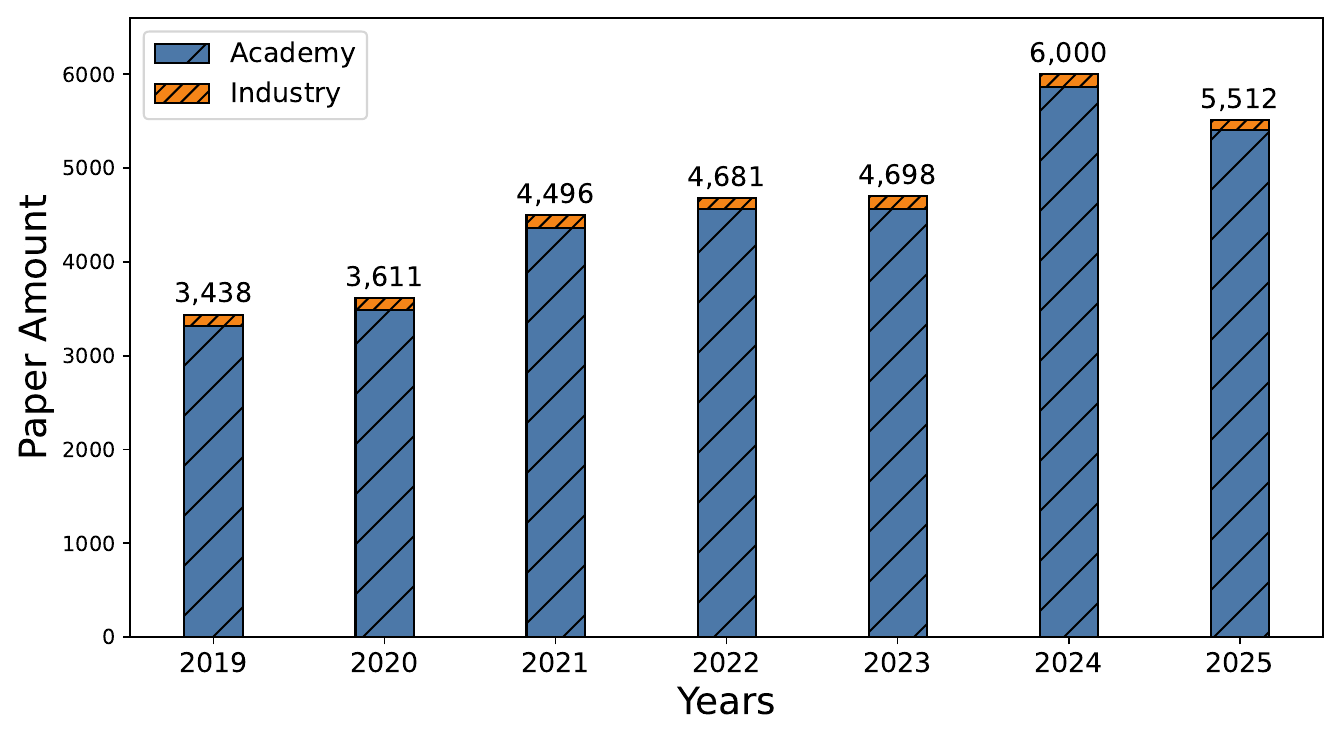}%
\label{fig:literature}}
\hfil
\subfloat[Country/Region]{\includegraphics[width=2.8in]{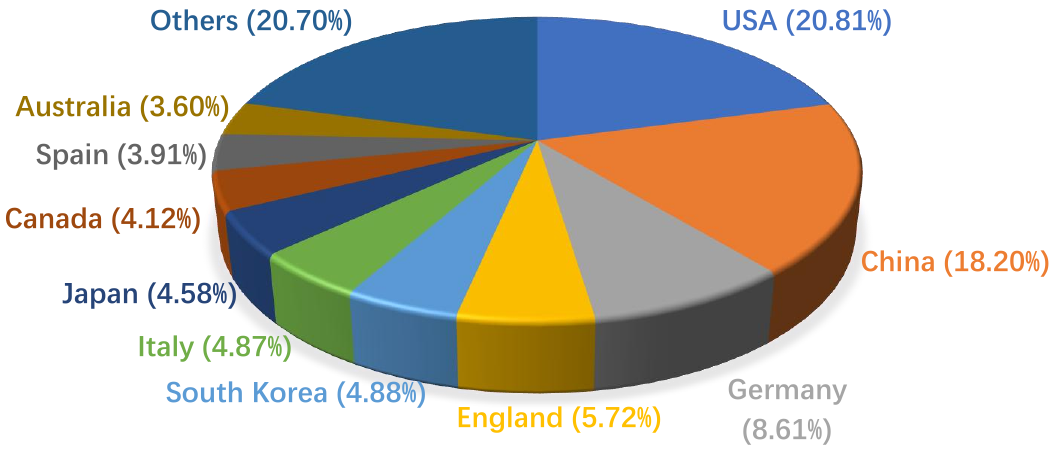}%
\label{fig:country}}
\caption{Literature distribution of VR/AR related research}
\label{fig_sim}
\end{figure*}

As a promising research area, VR/AR has been extensively investigated. Referring to the core database (SCIE, CPCI-S and ESCI) in Web of Science\footnote{Search keywords: VR or AR or virtual reality or augmented reality}, the annual VR/AR-related research literature has grown significantly in the last seven years.  As depicted in Fig.~\ref{fig_sim}\subref{fig:literature}, the volume of publications maintains a robust upward trend, showcasing active and deeply intertwined contributions from both academia and industry. Furthermore, this research interest is globally distributed, with the USA (20.81\%) and China (18.20\%) leading the contributions, followed closely by European and other Asian countries, as shown in Fig.~\ref{fig_sim}\ref{fig:country}. In the early days, research related to VR/AR focused mainly on VR/AR hardware (e.g., VR/AR devices). For example, they focused on (1) high-resolution, low-latency, wide-viewing angle display devices, including head-mounted displays (HMDs) and AR glasses; (2) high-precision head, hand, and eye tracking sensors; (3) devices that provide realistic haptic feedback, such as gloves and controllers. Moreover, some research focused on the fusion of VR/AR with other technologies, such as 5G and Internet of Things (IoT) technologies. Further, some research explored applications of VR/AR in fields such as education, entertainment, healthcare, and manufacturing. In short, the above research is mainly related to VR/AR hardware and the corresponding interaction technologies, as summarized in the development of VR/AR hardware-related research~\cite{Yushan2021, Abdullah2023}.

  

\begin{figure*}[h]
  \centering
  \includegraphics[width=0.85\linewidth]{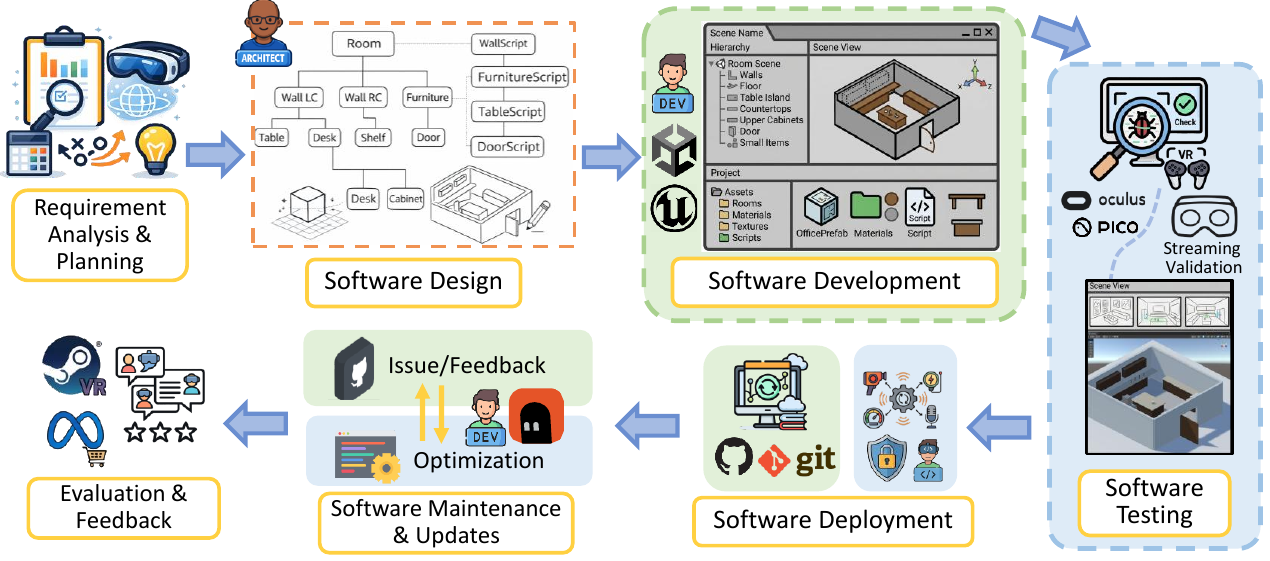}
  \caption{VR/AR Software Lifecycle}
  \label{lifecycle}
\end{figure*}

In addition to hardware-related issues, VR/AR software has also become a hot topic recently. As an example, the global VR/AR software market size was recorded at USD 46.11 billion by 2025, is anticipated to hit around USD 54.88 billion by 2026, and is expected to reach around USD 252.54 billion by 2035, at a CAGR of 18.54\% from 2026 to 2035~\cite{Zoting2026}. Driven by the rapid expansion of the VR/AR market, many VR/AR-related software applications have been released. Accordingly, research areas related to VR/AR software have received more attention. For example, they focus on the security and privacy mechanisms of VR/AR software, the graphics rendering of VR/AR software, the user experience of VR/AR software, and so on. We summarize and categorize the research related to VR/AR software according to the software life cycle, as shown in Fig.~\ref{lifecycle}. In particular, the VR/AR software life cycle consists of (1) Requirement Analysis and Planning; (2) Software Design; (3) Software Development; (4) Software Testing; (5) Software Deployment; (6) Software Maintenance and Updates; and (7) Software Evaluation and Feedback.

Although the impressive progress in studying VR/AR from a software perspective reveals its promising future, a multitude of existing studies from different organizations (e.g., academia and industry) and diverse disciplines (e.g., computer graphics, computer vision, software engineering, and system security) make it difficult for interested researchers to understand the state of the art in the VR/AR field. More importantly, VR/AR software has different characteristics from conventional software, e.g., a higher requirement for immersion and interactivity in 3D spaces. Correspondingly, the design and development of VR/AR software are more complex than conventional software. Moreover, VR/AR software has also caused new issues, especially related to immersive user experiences in virtual/realistic spaces. Consequently, many related empirical studies and software analysis/testing efforts have been made. Unfortunately, there is a lack of systematic work summarizing these existing studies about VR/AR software from the perspective of software engineering practice.

To bridge this gap, this paper aims to provide a systematic survey summarizing the state-of-the-art research on VR/AR software, along with a retrospection of VR/AR research from a software engineering practice perspective. Specifically, we will answer a number of research questions related to VR/AR software according to the software life cycle. We next briefly elaborate on these research questions as follows:
\begin{itemize}[leftmargin=*, labelindent=0pt,noitemsep,topsep=0pt]
\item \textbf{Requirement Analysis and Planning}: (1) How can we capture user needs and target user groups? (2) How can we determine the functions and features of a VR application? (3) How can we identify and assess technical feasibility and VR/AR market needs?
\item \textbf{Software Design}: (1) How can we design the user interface and user experience (UI/UX) for the VR environment? (2) What design principles can be applied to VR interactions?  (3) How can we create immersive scenes and content?
\item \textbf{Software Development}: (1) Which programming languages and frameworks are best suited for VR development? (2) How can we implement physical interaction and realism? (3) How can we handle compatibility issues with different VR platforms?
\item \textbf{Software Testing}: (1) How can we evaluate the performance and user experience of VR/AR applications? (2) What testing methods and tools can be applied to VR software? (3) How can we identify and fix VR-specific bugs?
\item \textbf{Software Deployment}: (1) How can we choose a suitable distribution VR/AR software platform? 
(2) How can we manage version control and software updates? (3) How can we ensure user privacy and data security?
\item \textbf{Software Maintenance and Updates} (1) How can we continuously improve VR applications based on user feedback? (2) How can we handle the adaptation of new technologies and platforms? (3) How can we maintain user engagement and interest?
\item \textbf{Evaluation \& Feedback}: (1) How can we collect user usage data to evaluate the success of the app? (2) How can we conduct user satisfaction surveys and experience studies? (3) How can we translate evaluation results into guidance for future development?
\end{itemize}

Therefore, the researchers can have a thorough understanding of the advantages and limitations of the existing software analysis techniques in VR/AR. Based on our analysis, we point out the current challenges and suggest possible future directions for VR/AR from a software-related perspective. Overall, our work provides a comprehensive review of the current progress of the VR/AR software community, enabling researchers to obtain an overview of this thriving field and make progress toward advanced practices. The contributions of this paper are concluded as follows:
\begin{itemize}[leftmargin=*, labelindent=0pt,noitemsep,topsep=0pt]
\item We collect more than 110 relevant papers about VR/AR software from the top-tier conferences as well as journals in software engineering, security, and VR/AR.
\item According to the VR/AR software life cycle, we present a categorization of these studies and conduct a detailed analysis of publication trends, distribution of publication venues, insights, findings, and so on.
\item We summarize research goals and technical contributions according to each stage of the VR/AR software life cycle.
\item We pinpoint key open research challenges of the VR/AR software life cycle and provide insightful guidelines on the software practices of VR/AR.
\end{itemize}

\begin{figure*}[h]
  \centering
  \includegraphics[width=0.8\linewidth]{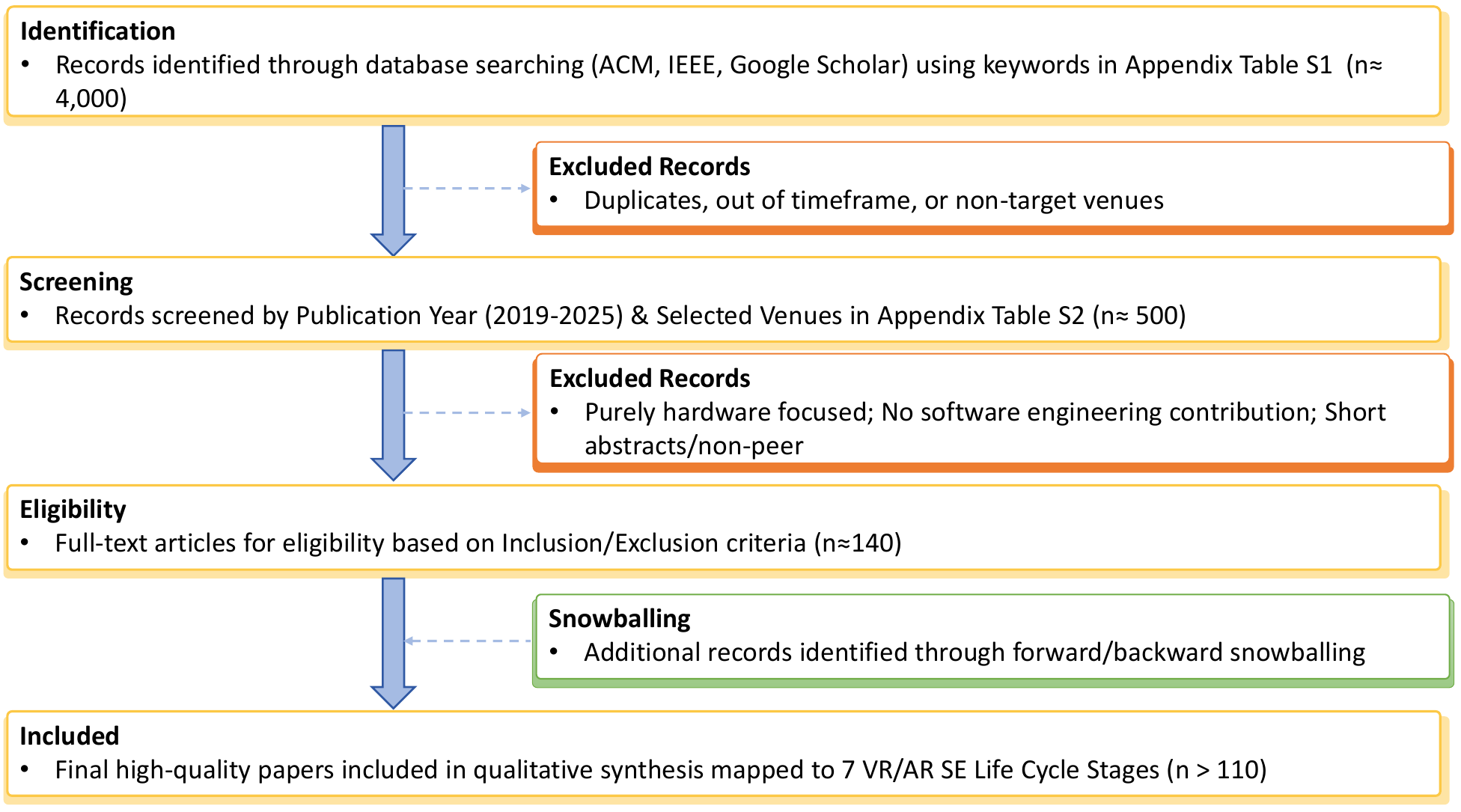}
  \caption{Literature collection process of VR/AR papers}
  \label{search_process}
\end{figure*}

\subsection{Comparison with Existing Surveys}
While there is a growing body of literature on VR/AR, a comprehensive review from a holistic perspective of software engineering practices remains notably absent. Existing survey papers can be broadly classified into three categories, each leaving a critical gap:

\emph{(1) Hardware and Application-Centric Surveys}: The majority of reviews heavily focus on hardware advancements or specific vertical applications (e.g., education, healthcare)~\cite{Abdullah2023, kaminska2022virtual}. While valuable, they treat the underlying software as a black box, overlooking complex engineering practices.

\emph{(2) Domain-Specific Software Surveys}: Other reviews address isolated software issues, such as cybersickness evaluation~\cite{Rebenitsch2016} or metaverse security threats~\cite{Pınar2022, Wang2023}. Although deeply insightful in their niches, they lack a full software life-cycle perspective, thereby failing to capture some important software engineering practices, e.g., how do requirement decisions affect downstream UI testing or deployment?

\emph{(3) Traditional Software Engineering Surveys}: Despite numerous software engineering surveys for web~\cite{li2025_1}, mobile~\cite{yu2025vision}, and AI applications~\cite{martinez2022software}, VR/AR software fundamentally diverges due to 6DoF spatial interactions, stereoscopic rendering, and the stringent demands of immersive Quality of Experience (QoE).

To the best of our knowledge, \textit{this is the first systematic survey that bridges VR/AR technologies and comprehensive software engineering practices} across a complete seven-stage VR/AR software life cycle.

\subsection{Key Findings}
By synthesizing the state-of-the-art literature, we distill several key findings that define the current landscape of VR/AR software engineering practices:
\begin{itemize}[leftmargin=*, labelindent=0pt,noitemsep,topsep=0pt]

\item \emph{Fundamental Paradigm Shift in Design and Testing}: VR/AR development requires a complete departure from 2D metaphors. VR/AR software design heavily relies on 3D egocentric/exocentric techniques and privacy-driven adaptations, while testing demands spatial interaction automation and autonomous scene exploration rather than traditional static GUI analysis/testing.
\item \emph{Ubiquitous Impact of Generative AI and LLMs}: AI is fundamentally reshaping every phase of the VR/AR life cycle. From LLM-assisted requirement parsing and text-to-3D rapid prototyping to AI-driven immersive authoring (e.g., in-situ voice coding) and multi-agent cybersickness assessment, generative AI and other AI technologies are transitioning VR/AR engineering toward intelligent co-creation.
\item \emph{Severe and Unique Deployment Vulnerabilities}: The deployment phase faces critical, domain-specific security threats. Spatial UI spoofing, hidden operations, and side-channel attacks via motion tracking emphasize the urgent need for robust, VR-native security-by-design frameworks and longitudinal firmware auditing.
\item \emph{High Coupling in Maintenance}: Fixing VR-specific performance bugs and ``bad smells'' is uniquely complex, frequently requiring coordinated modifications across source code, 3D assets, and engine-specific configuration files.
\end{itemize}

\section{Research Methodology and Literature Collection}
To ensure a rigorous, transparent, and reproducible review process, we followed a systematic literature review methodology inspired by the Preferred Reporting Items for Systematic Reviews and Meta-Analyses (PRISMA) guidelines~\cite{Helbach2023}. The literature collection and filtering process is meticulously designed to capture high-quality studies at the intersection of Software Engineering and VR/AR.

\subsection{Search Strategy and String Formulation}
Our search strategy was executed across leading academic databases, primarily the ACM Digital Library, IEEE Xplore, and Google Scholar. To construct comprehensive search strings, we combined two main conceptual domains: the Software Life Cycle Domain and the Immersive Technology Domain. Specifically, we extracted representative keywords for all seven stages of the software life cycle (e.g., ``requirement analysis'', ``UI/UX'', ``bug detection'', ``version control''). These were then combined using Boolean operators (AND) with technology-specific terms (e.g., ``Virtual Reality'', ``Augmented Reality'', ``VR'', ``AR''). We further elaborate on the detailed search strings formulated for each life cycle stage in Appendix presented in the supplemental material (Table~S1).

\subsection{Venue Selection and Quality Assessment}
To guarantee the academic rigor and practical relevance of the surveyed literature, we strategically restricted our primary search to top-tier venues (Core A/A* or CCF A/B ranking), including premier conferences and journals across four intersecting domains: Software Engineering (e.g., ICSE, FSE, ASE, TSE, TOSEM), Security \& Privacy (e.g., S\&P, CCS, USENIX Security, NDSS), Human-Computer Interaction (e.g., CHI, UbiComp), and specialized VR/AR venues (e.g., IEEE VR). A complete list of the targeted venues, along with our quality assurance rationale based on academic rankings, is provided in Appendix III (Table~S2).

\subsection{Inclusion and Exclusion Criteria (Screening Process)}
The retrieved literature underwent a rigorous multi-stage screening process, which is conceptually mapped in Fig.~\ref{search_process}. The filtering was strictly governed by the following criteria:
\begin{itemize}[leftmargin=*, labelindent=0pt,noitemsep,topsep=0pt]
\item \emph{Inclusion Criteria}: (1) The study should explicitly address a software engineering problem (e.g., requirement engineering, design, testing, maintenance, and security) within a VR/AR context. (2) The study should reflect recent progress (i.e., published between 2019 and 2025), capturing the most recent technological advancements. (3) The study must be a full-length, peer-reviewed article.
\item \emph{Exclusion Criteria}: (1) Papers purely focused on hardware improvements (e.g., optics, battery life) without software-level implications. (2) Studies utilizing VR/AR solely as a demonstration tool for non-CS research (e.g., clinical psychology experiments) without contributing to software engineering practices. (3) Posters, extended abstracts, or non-peer-reviewed preprints (unless exceptionally highly cited or high-quality ones).
\end{itemize}

\subsection{Data Extraction and Final Distribution}
In the final stage, two authors independently conducted a full-text review of the screened papers to extract key contributions, methodologies, and findings. Discrepancies were resolved through discussion with a third senior author. Additionally, a backward and forward snowballing technique was applied to the reference lists of key papers to capture any highly impactful literature missed by the automated search.

Following this rigorous protocol, we finally curated and analyzed about 110 high-quality papers. As categorized in Table~\ref{distribution}, the distribution of papers across the life cycle reveals that Software Deployment and Software Testing currently dominate the research landscape, while Evaluation \& Feedback remains an underexplored frontier.

\begin{table*}
  \caption{Paper distribution according to VR/AR Software Life Cycle}
  \label{distribution}
  \resizebox{0.95\linewidth}{!}{
  \begin{tabular}{cccccccc}
    \toprule
    Lifecycle & \tabincell{c}{{Requirement Analysis}\\ {\& Planning}}  &\tabincell{c}{{Software}\\ {Design}}&\tabincell{c}{{Software}\\ {Development}}&\tabincell{c}{{Software}\\ {Testing}}&\tabincell{c}{{Software}\\ {Deployment}}&\tabincell{c}{{Software Maintenance}\\ {\& Updates}}&\tabincell{c}{{Evaluation}\\ {\& Feedback}}\\
    \midrule
    \# of Papers & 18 & 11& 16& 20& 35 & 6 & 4\\
  \bottomrule
\end{tabular}}
\end{table*}

\section{VR/AR Requirement Engineering and Planning}

Although VR/AR applications share some common characteristics with mobile applications, they are fundamentally different. Running on top of the underlying VR/AR hardware, VR/AR applications provide users with an immersive experience in a virtual 3D space. Moreover, 3D virtual objects can be overlaid with real-world environments to achieve the mixed reality (MR) effect. Additionally, users can use controllers, haptic gloves, virtual keyboards, and hand gestures to control objects in the VR/AR 3D space. In contrast, mobile applications are typically restricted to 2D touchable screens limited by mobile devices' form factors and operational manners.  

\begin{figure}[h]
  \centering
  \includegraphics[width=\linewidth]{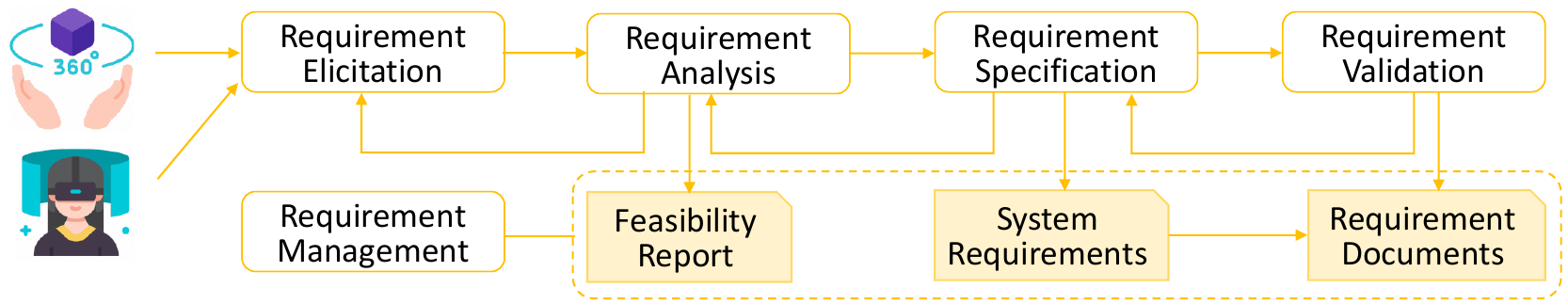}
  \caption{Requirement engineering for developing VR/AR applications}
  \label{fig:re-overview}
\end{figure}

These unique features of VR/AR systems not only diversify the VR/AR software ecosystem but also pose new challenges in adopting VR/AR in various industrial domains. Conducting a comprehensive requirement engineering (RE) is the first yet crucial step before developing and deploying VR/AR software in an industrial domain~\cite{10.1145/3649595}. As shown in Fig.~\ref{fig:re-overview}, a typical requirement engineering process of VR/AR software is composed of 1) requirement elicitation, 2) requirement analysis, 3) requirement specification, 4) requirement validation, and 5) requirement management~\cite{lamsweerde2009requirements}. Next, we briefly elaborate on the research issues of VR/AR across the entire RE process, according to the five phases mentioned above.

\subsection{Requirement Elicitation}

Requirement elicitation (or requirement collection) is essentially a process of acquiring context knowledge of the potential usage domain, where context knowledge mainly includes stakeholders' backgrounds and domains, project objectives, actors, resources, and problems. Like most other software systems, it is also necessary to conduct a comprehensive requirement elicitation for VR/AR applications before the VR/AR development. Several requirement elicitation approaches are summarized in a recent study~\cite{10.1145/3578527.3578536}. Generally, \emph{direct inspection} has widely been adopted in diverse VR/AR industrial domains, such as automotive, construction, education, entertainment, gaming, and healthcare. In a nutshell, direct inspection can determine the requirements based on the potential interests of stakeholders. Examples include a posture-adaptive VR system based on haptic gloves~\cite{10.1145/3631453} and automotive mechanics simulations~\cite{quevedo2017virtual}. Despite the simplicity of direct inspection, acquiring domain knowledge is crucial, especially for those VR/AR applications for healthcare and industry domains. 

Besides direct inspection, other requirement elicitation approaches include questionnaires, interviews, user case studies, and prototypes similar to conventional software (e.g., mobile applications). Different from conventional software systems, VR/AR requirement elicitation may involve more immersive interactions with potential users in VR/AR environments. For example, comprehensive feedback data of fingertips and finger phalanges provided by haptic gloves were collected to further improve the VR users' perception performance~\cite{10.1145/3678567}. Additionally, questionnaires based on the developed VR/AR prototype can be employed and analyzed to enhance the system in future development~\cite{li2024gamification}. Notably, the adoption of VR/AR technologies can capture users' requirements even in developing conventional software, as mentioned in~\cite{9974443}.

\subsection{Requirement Analysis}
Similar to conventional software applications, VR/AR applications have adopted similar requirement analysis tools~\cite{10.1145/3649595}. Typical requirement analysis approaches include goal-oriented schemes, object-oriented methods, flow chart diagrams, etc.~\cite{lamsweerde2009requirements}. In goal-oriented schemes, goals can be roughly categorized into behavioral goals and soft goals. A behavioral goal is either \emph{achievable} or \emph{avoidable}, where an achievable goal prescribes an intended behavior and an avoidable goal depicts a goal to be avoided. For example, drivers' interests can be coordinated with the operational goals of automatic vehicles by properly analyzing achievable and avoidable goals with consideration of conflict-handling strategies in AR-enabled driving scenarios~\cite{10.1145/3678511}. 

In addition to goal-oriented schemes, object-oriented methods have been typically adopted in VR/AR applications. Among these object-oriented methods, Unified Modeling Language (UML) has been widely adopted in modern software development. Without exception, UML has also been widely adopted in the development of VR/AR applications. Karre et al.~\cite{10.1145/3551349.3560514} proposed a role-based meta-modeling language based on UML for VR/AR software development. Moreover, a web-based modeling approach was proposed to model AR applications~\cite{10.1145/3652620.3687779}. Additionally, a UML-based 3D collaborative system was developed to capture users' requirements~\cite{10.1145/3274298}. Notably, conventional software engineering techniques, such as the detection of requirement conflicts and risk analysis have also been adopted in this phase. Furthermore, with the rapid development of LLMs, researchers are exploring the usage of natural language processing (NLP) techniques to automatically parse unstructured stakeholder interviews and translate them into structured VR requirement templates, significantly reducing manual effort.

\subsection{Requirement Specification} 

After requirement analysis, the requirement specification is obtained, though this phase is also intertwined with previous phases, thereby forming the RE lifecycle. 
In general, requirement specification takes a set of agreed statements as inputs, including objectives, system requirements, running environments (e.g., Oculus VR HMDs or Apple Vision Pro), and domain properties. It then outputs requirement documents with detailed specifications. Common description approaches for requirement specifications include unrestricted natural language, structured natural language, diverse diagrams, and formal specifications~\cite{lamsweerde2009requirements}, where formal specifications may adopt propositional logic, first-order predicate logic techniques, etc. Regarding requirement specification templates, the latest international standard is ISO/IEC/IEEE 29148~\cite{ISO29148}, which weeds out IEEE Recommended Practice for Software Requirements Specifications (IEEE 830-1998)~\cite{IEEE830-1998}. 
Recently, to address the unique challenges of VR software, such as multi-modality and platform fragmentation, specific requirement specification tools have been proposed. For instance, based on their previous role-based meta-model~\cite{10.1145/3551349.3560514}, Karre et al.~\cite{Karre2025} introduced VReqST, a web-based requirement specification tool tailored for VR products. Instead of relying solely on plain natural language, it utilizes structured, JSON-based model templates and an underlying validator engine to specify precise VR requirements. This approach ensures the logical correctness of VR-specific attributes and supports specification versioning, thereby significantly reducing rework loads and bridging the communication gap among stakeholders.

\subsection{Requirement Validation and Management}

Requirement validation (RV) is a necessary step to guarantee the quality of requirement specifications. During the RV process, the requirements specified by requirement documents will be checked in terms of validity, consistency, completeness, and verifiability. Typical RV approaches include inspections, reviews, simulations, and formal checks. For example, reviewers may check for errors or inconsistencies in the requirement documents. Further, a runnable VR/AR system prototype is often adopted to demonstrate the system's feasibility to end-users or customers. Recent studies, such as~\cite{10.1145/3290605.3300826} have demonstrated the feasibility of developing a prototype of VR apps based on pre-defined templates. 

\begin{figure}[h]
  \centering
  \includegraphics[width=\linewidth]{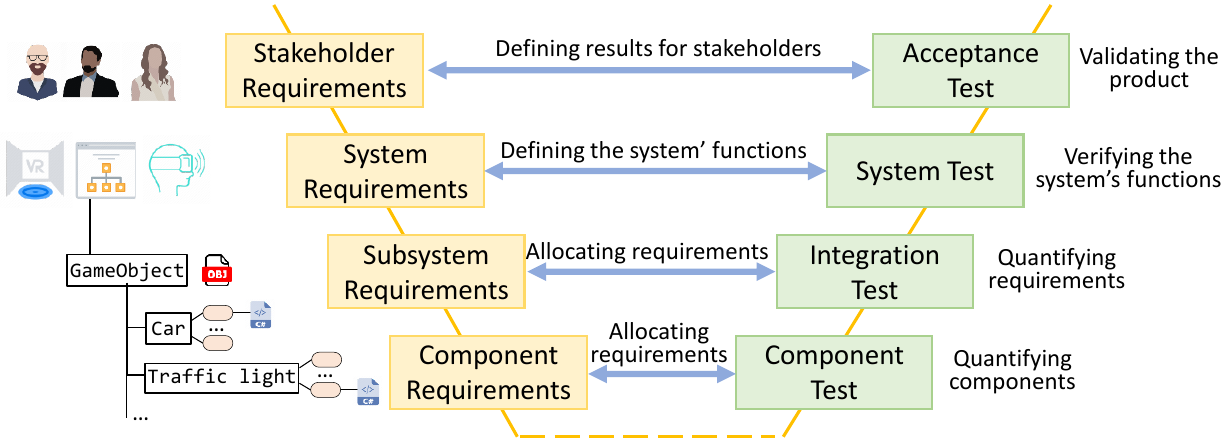}
  \caption{Requirement management of VR/AR applications in the V model}
  \label{fig:v-model}
\end{figure}

The requirements of modern software systems are often undergoing a change. As a result, the requirements have to evolve to respond to the changed requirements raised by stakeholders. Maintaining the changed requirements during the entire RE process is a necessity to guarantee the developed software systems. Many different tools have been developed for requirement management. Among them, the V model is a typical approach. Fig.~\ref{fig:v-model} depicts the V model adopted for requirement management of VR/AR applications. Take a VR-based traffic flow simulation system as an example~\cite{HASAN2021102985}, in which the stakeholders express their demands in visualizing microscopic traffic in a city and providing the available optimized routes. The system engineers may offer a suite of solutions by defining the system's functions while the developers will design and implement the solutions with consideration of physical constraints. This fully traceable RE process will keep evolving with multi-round analysis and validation. 

\summary{\revise{Requirement engineering is a crucial step to capture stakeholders' needs before developing VR/AR applications. Unlike 2D mobile or desktop applications, VR/AR poses unique challenges in analyzing multi-modal interactions. While most VR/AR applications still adopt canonical approaches, specialized structured templates (e.g., JSON-based models) are emerging. Furthermore, a promising insight is the integration of LLMs to automatically parse unstructured stakeholder interviews and translate them into these precise VR/AR requirement specifications, significantly reducing manual effort and communication gaps.}}

\section{VR/AR Software Design}
Software design is an interactive process, through which requirements are translated into a ``blueprint'' for constructing the software. Today, we have well-established software design methodologies for creating experiences on computers or mobile devices (e.g., smartphones and tablets)~\cite{6312969,zhu2005software,stephens2015beginning,ghezzi1991fundamentals}. When designing traditional software, contributing design ideas has a relatively low barrier. For example, designers draw wireframes (e.g., rectangles for the device screen and sketch UI elements with boxes and lines) to create UI layouts~\cite{hamm2014wireframing} and employ UML models (e.g., activity diagrams and sequence diagrams) to define potential events~\cite{gomaa2011software}.

However, this traditional 2D design process begins to break down when applied to VR/AR (3D) software design. For most design participants, sketching in 3D can be challenging, and relying solely on 2D tools such as pen and paper or whiteboards often limits ideas to those dimensions. Additionally, 3D design tools require a high level of skills to quickly realize concepts. Furthermore, there is a vast array of novel interaction techniques (e.g., world-in-miniature, virtual hand) in VR/AR, making it difficult for team members without a background in 3D development to contribute effectively.

\begin{figure*}[t]
  \centering
  \includegraphics[width=0.85\linewidth]{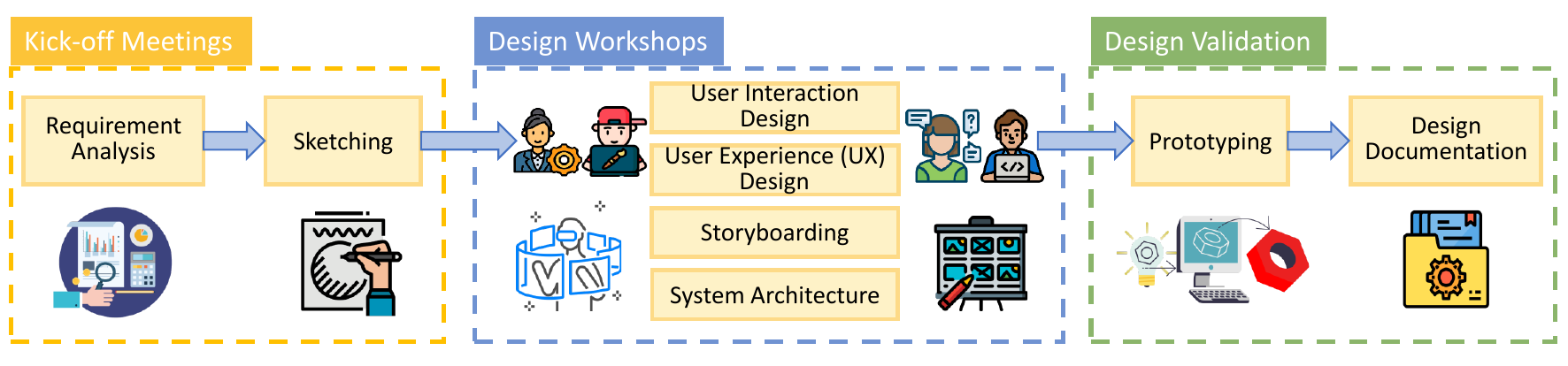}
  \caption{Pipeline of VR/AR software design}
  \label{fig:design_pipeline}
\end{figure*}

Existing research on VR/AR software design has primarily focused on UI design~\cite{10.1145/3491102.3517714,10.1145/3489849.3489876}. However, all roles within the project team should participate in the design process, including concept developers, interaction designers, content authors, and technical developers~\cite{krauss2021current}. In the following, we will provide a comprehensive review of the VR/AR software design process. Fig.~\ref{fig:design_pipeline} illustrates the pipeline of the VR/AR software design process, and we will further explain the key steps as follows.

\subsection{Sketching}
\begin{figure}[h]
  \centering
  \includegraphics[width=1\linewidth]{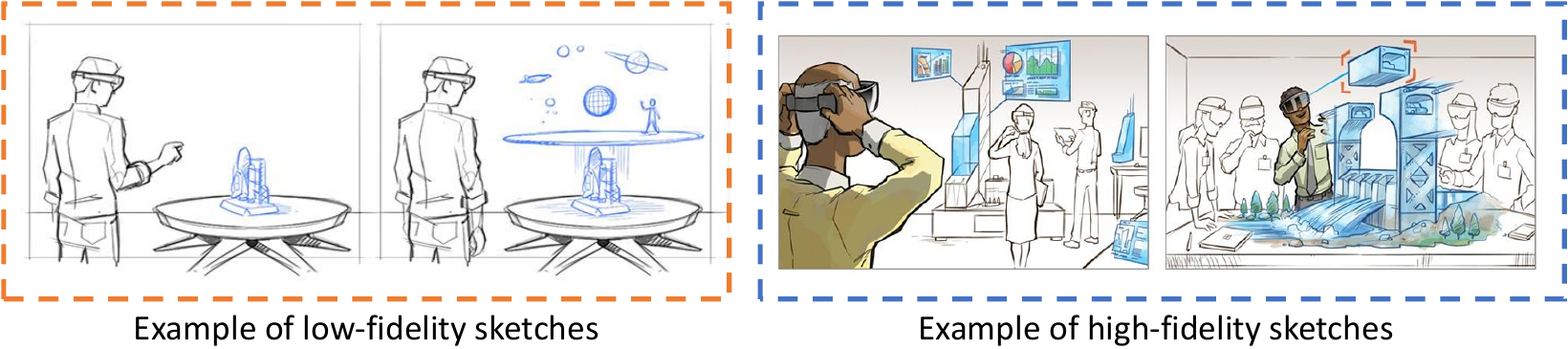}
  \caption{Examples of different fidelity sketches for VR/AR software design~\cite{microsoft2022mixedreality}.}
  \label{fig:sketches}
\end{figure}

At the beginning of the software design stage, compared to using cumbersome interactive user interface building tools directly, designers prefer to develop early ideas through hand-drawing or employing other ``low-fidelity techniques'' ~\cite{rettig1994prototyping}. This approach is favored because the designer's goal at this stage is to develop the overall layout and structure of components rather than to refine the detailed look-and-feel~\cite{landay1995interactive}. An efficient sketch should adhere to the following requirements:
1) It should be quick to create, have a low level of detail, and be ambiguous to encourage discussion. 
2) Software engineers can utilize sketches to achieve a common understanding and avoid strong formalization, which is often seen as a hindrance~\cite{buxton2010sketching, shipman1999formality}. Fig.~\ref{fig:sketches} shows examples of low-fidelity and high-fidelity sketches for VR/AR software design.

To overcome the dimensional limitations of sketches that rely on pen and paper while preserving the benefits of being quick to create and avoid strong formalization, several approaches have been proposed for VR/AR software design. Hagbi et al.~\cite{hagbi2010place} define a visual language that enables In-Place Sketching. It supports generating virtual content from hand sketches through 3D augmentation techniques. Langlotz et al.~\cite{langlotz2012sketching} implement an authoring tool for mobile AR applications that allows users without 3D modeling experience to freehand sketch 3D primitives in a simple way, enabling users to create new 2D and 3D objects as well as replicate existing 3D objects. Gasques et al.~\cite{gasques2019you} develop an interactive sketching tool, PintAR, which enables the authoring and rapid prototyping of situated experiences by allowing designers to conceptualize using a digital pen for sketching and a head-mounted display for visualizing and interacting with virtual content. Moreover, Buruk et al.~\cite{10.1145/3464327.3464330} propose Immersive Video Sketching (IVS), a low-cost prototyping method for early-phase VR/AR design that can be easily employed by novice and non-expert designers. Drey et al.~\cite{10.1145/3313831.3376628} introduce VRSketchIn, a VR sketching application that utilizes a 6 degree-of-freedom (6DoF) tracked pen and a 6DoF-tracked tablet as input devices to enhance perception and understanding of 3D volumes, which is particularly beneficial for VR/AR design. 
Beyond sketching in virtual space, Lolambean et al. from Microsoft encourage using body-storming to build preliminary ideas (i.e., building physical props using simple, inexpensive materials to represent virtual 3D objects, user interfaces, and animations)~\cite{microsoft2022mixedreality}.

\subsection{User Interaction Design}
Before the rise of VR/AR, most software UI design only considered a set of two-dimensional UI components, including (i) \emph{input devices}, such as the mouse and touchscreen, (ii) \emph{output devices}, such as mobile phones and tablets, (iii) \emph{interaction techniques} like drag-and-drop and pinch to zoom, and (iv) \emph{UI metaphors}, such as menus, sliders, and gestures~\cite{3DUIlaviola20173d}. However, these UI design techniques are often inappropriate for contemporary VR/AR technologies. Modern VR/AR often provides users with an immersive experience through HMDs (also known as headsets). By wearing these headsets, users enter a three-dimensional user interface that allows for 6DoF manipulation of virtual 3D objects. Most VR/AR headsets are also equipped with controllers, which serve dual purposes: (1) they act as a means of user input, and (2) they provide output, allowing users to experience haptic feedback through vibrations and other sensations~\cite{9152604}. These VR/AR devices achieve a higher level of immersion compared with other mobile devices or computers.

\begin{figure}[t]
  \centering
  \includegraphics[width=\linewidth]{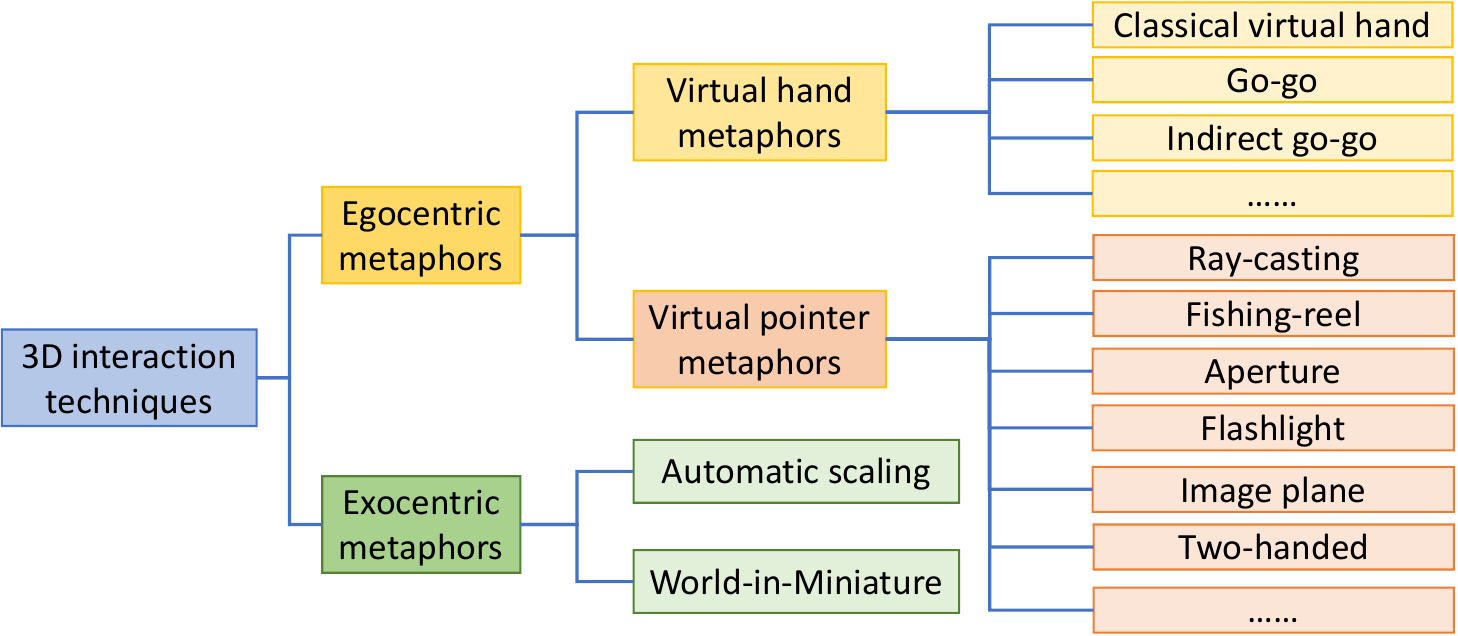}
  \caption{Common 3D interaction techniques for VR/AR~\cite{3DUIlaviola20173d}}
  \label{fig:3D_interaction_technique}
\end{figure}

\begin{figure}[h]
  \centering
  \includegraphics[width=1\linewidth]{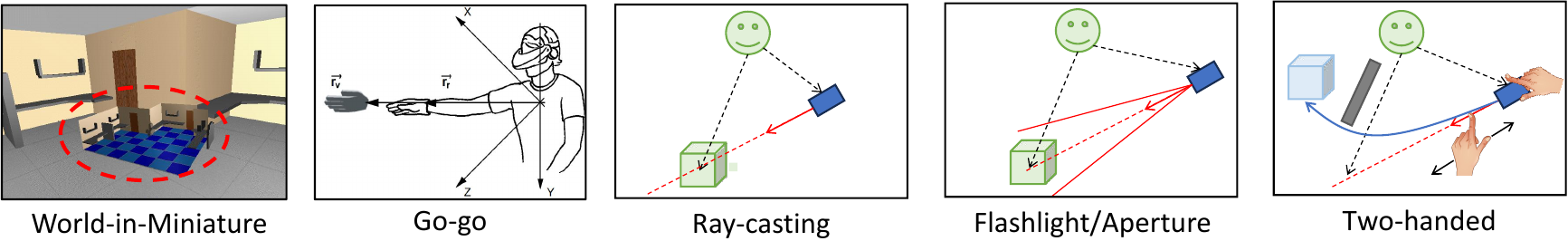}
  \caption{Examples of Common 3D interaction techniques~\cite{world-in-miniature}}
  \label{fig:3D_interaction_technique_examples}
\end{figure}

Since VR/AR software experience differs from other software in terms of \emph{dimensionality}, \emph{interaction}, and \emph{immersion}, new interaction techniques should be employed. The 3D VR/AR user interaction techniques can be classified by metaphors into exocentric metaphors and egocentric metaphors, with the latter further divided into virtual hand metaphors and virtual pointer metaphors~\cite{3DUIlaviola20173d}. Fig.~\ref{fig:3D_interaction_technique} illustrates common 3D interaction techniques for VR/AR, and Fig.~\ref{fig:3D_interaction_technique_examples} shows examples of these common 3D interaction techniques. 

When designing user interactions for VR/AR, interaction designers should choose the most suitable interaction techniques based on the diverse requirements, considering parameters associated with specific manipulation tasks (e.g., task space, object size, distance from objects to the user, and the physical and psychological states of the user, etc.)~\cite{3DUIlaviola20173d}. Additionally, privacy constraints are increasingly shaping VR/AR interaction design. Since immersive systems rely heavily on continuous sensing (e.g., spatial mapping, audio capture), Rajaram et al.~\cite{Rajaram2025} proposed a design space and a visualization tool for privacy-driven adaptation techniques, helping designers gracefully degrade or substitute interaction modalities when users restrict certain sensor permissions.

\subsection{Storyboarding}
\begin{figure}[t]
  \centering
  \includegraphics[width=1\linewidth]{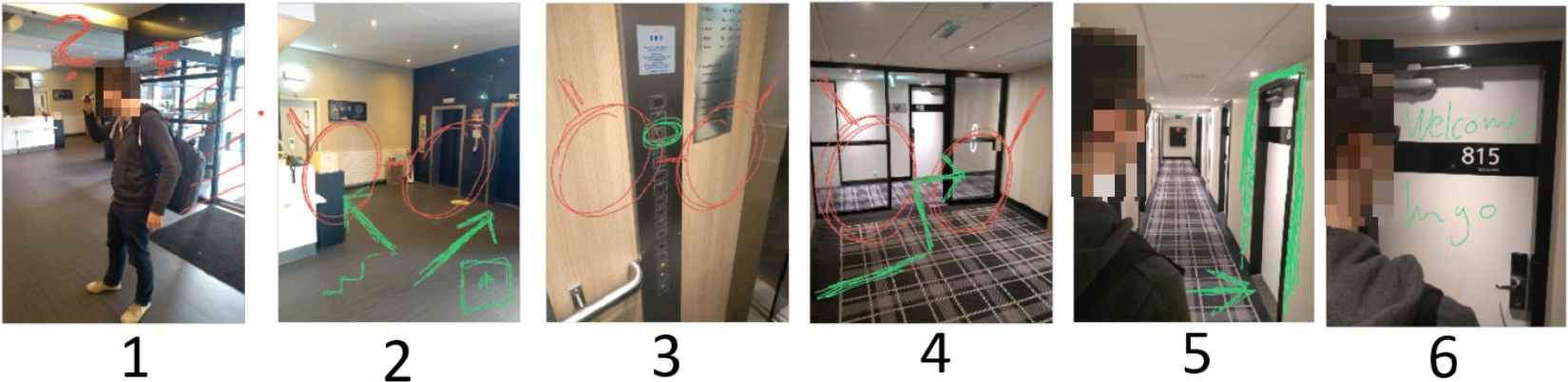}
  \caption{Storyboard for an AR indoor navigation app~\cite{borsting2022software}}
  \label{fig:storyboarding_example}
\end{figure}

Determining the most effective way to communicate proposed ideas and concepts depends on the target audience. When introducing new ideas to team members, low-fidelity ``sketches'' are often sufficient to provide a quick understanding. However, when presenting ideas and concepts to new stakeholders or potential users, storyboards are a more effective and engaging approach. Additionally, storyboards have the advantage of clearly illustrating a sequence of activities required for an event. Fig.~\ref{fig:storyboarding_example} shows an example of storyboarding design in AR application development. The storyboard is a graphical representation of a narrative originating from film and TV productions~\cite{hart1998art}. They consist of a chronological series of images that follow a sequence of events, accompanied by annotations to visualize common interaction possibilities.

Similar to storyboarding in traditional software design, VR/AR software design can also adopt classical 2D storyboard shots. 2D storyboards can be accomplished using both sketches and pictures. Users with good drawing skills often prefer sketch-based methods, as they allow for more freedom of expression and enable iterative improvements towards higher fidelity~\cite{10.1007/978-3-031-55722-4_13}. Some development teams may include a storyboard artist who captures difficult-to-describe ideas, sketches them out quickly during meetings, and then expands sketches into higher-fidelity versions~\cite{microsoft2022mixedreality}.
 
 Within the context of AR software design, picture-based storyboarding is often more popular than sketch-based storyboarding. This preference arises because real-world images can effectively represent the physical environment, while virtual objects can be illustrated through drawings or stickers. To distinguish between virtual objects and the physical world, it is a good practice to use black and white photos to represent the physical environment while employing different colors to highlight virtual objects, movements, and interactions~\cite{borsting2022software}.

Compared to sketching, storyboarding conveys the overall flow and aesthetic appearance of the experience efficiently, but it requires more effort. On one hand, more pictures or high-fidelity drawings need to be produced, which may sometimes require the assistance of an artist. On the other hand, different user travel routes can result in varying sequences of events, leading to multiple versions of storyboards. In practice, high-fidelity storyboards can not only effectively communicate ideas to potential users or new stakeholders, but also facilitate precise communication of concepts with asset artists~\cite{microsoft2022mixedreality}. Recently, generative AI has begun to reshape VR/AR software design. Techniques such as Text-to-3D and generative 3D Gaussian splatting allow designers to rapidly generate 3D assets, skyboxes, and environmental prototypes simply through natural language prompts or 2D sketches. This AI-driven design paradigm significantly lowers the barrier to entry, enabling rapid prototyping without requiring extensive 3D modeling expertise.

\subsection{User Experience (UX) Design}
Contemporary VR/AR headsets can provide an immersive virtual experience by enveloping the senses with computer-generated stimuli. Compared to traditional 2D user interfaces, the immersive and interactive nature of VR/AR offers users a completely different experience. However, unfamiliar user interface elements, physical discomfort, and user confusion pose significant challenges for designers. We will discuss VR/AR user experience design concerns in terms of user comfort, spatial awareness, and haptic feedback as follows.

\subsubsection{User Comfort}
Unlike traditional 2D UX design, the greatest challenge in VR/AR lies in user comfort. Low-quality VR/AR applications can lead to user discomfort and even have adverse health effects, severely dampening users' enthusiasm for using VR/AR. Cybersickness is one of the most common physical discomforts reported by users~\cite{Rebenitsch2016}. VR/AR headsets utilize stereoscopic 3D (S3D)~\cite{penn2018making, elmqaddem2019augmented}, which provides two distinct two-dimensional (2D) images for each eye of the users. The user's brain then constructs a corresponding stereoscopic 3D scene, creating an illusion of depth based on these two images. Improper rendering can cause cybersickness in VR/AR, with symptoms such as headaches, disorientation, and nausea, which can potentially affect users' health and safety, hindering the development and growth of VR applications~\cite{shih20203d,munafo2017virtual,laviola2000discussion,peng2020walkingvibe,nichols2002health}.

To mitigate cybersickness, many UX practices have been proposed. Firstly, one of the most common practices is to move in the 3D virtual space without the physical movement of users. For instance, both remaining stationary and moving through teleportation can make the motion implicit, allowing users to feel comfortable~\cite{Clifton2020,laviola2000discussion}. 
Secondly, it is advised not to accelerate, rotate, or decelerate the VR camera since the vestibular system can feel acceleration and rotation but not fixed-speed linear motion~\cite{Pruett2017}. 
Thirdly, a technique known as ``temporary change to blindness'' is employed to avoid shots that may induce cybersickness~\cite{Pruett2017}. The idea is based on the brain's tendency to fill in gaps in our perceptual streams (e.g., we don’t notice it when we blink). A temporary change to blindness is used to omit visual information that may cause discomfort by simply cutting to black or instantaneously changing the viewpoint. For example, when a person needs to ascend a spiral staircase, experiencing the entire process involves significant non-uniform head rotation, which could induce cybersickness in many individuals. Simply cutting from the moment they step onto the first few steps to the point, at which they have already reached the top, with a brief black interstitial, feels both more natural and expected.

\subsubsection{Spatial Awareness}
VR/AR provides users with a 3D virtual space to explore. However, helping users understand unfamiliar 3D environments is a challenge. In the VR/AR context, \textit{spatial awareness} involves how users navigate, interact with, and relate to the virtual or augmented environment. Effective spatial awareness enables users to intuitively comprehend their surroundings, leading to a more immersive and engaging experience. Conversely, inadequate spatial awareness can lead to confusion, disorientation, and even cybersickness.

When users enter a brand-new 3D virtual space, it is essential to help them establish 3D spatial knowledge. Initially, users are only capable of acquiring egocentric information, such as landmarks and procedural knowledge (route knowledge)~\cite{3DUIlaviola20173d}. However, a higher level of spatial knowledge, known as survey knowledge, takes a significantly longer time to develop. 
To facilitate the acquisition of exocentric survey knowledge, allowing users to explore the environment is a common practice. However, freely exploring complex environments can quickly result in disorientation. As a result, developers often integrate both natural and artificial cues to help users navigate within environments effectively~\cite{3DUIlaviola20173d,stefanucci2022effects}. Natural cues consist of lighting and shadows that offer directional and depth information, special illumination that emphasizes landmarks, and open spaces (such as doors in walls) that indicate preferred directions for user movement. In contrast, artificial cues include (i) maps providing spatial orientation, (ii) signs indicating routes, (iii) reference objects assisting users in estimating size and distance (e.g., familiar objects such as coins, matchboxes, and humans), and (iv) trails helping users retrace their steps~\cite{3DUIlaviola20173d}.

\subsubsection{Haptic Feedback}
Despite the high level of visual immersion, genuine haptic feedback remains unattainable for VR/AR. Most controllers for VR/AR headsets on the market can only provide basic haptic feedback through vibrations of varying frequencies and intensities~\cite{MetaHorizonOS}. As a result, developers often avoid interactions that require difficult-to-implement haptic feedback. For instance, lifting a heavy box or grasping a sharp object may not align with users' expectations, as the rigid constraints they anticipate cannot be realized~\cite{Pruett2017}.

\begin{figure}[h]
  \centering
  \includegraphics[width=0.8\linewidth]{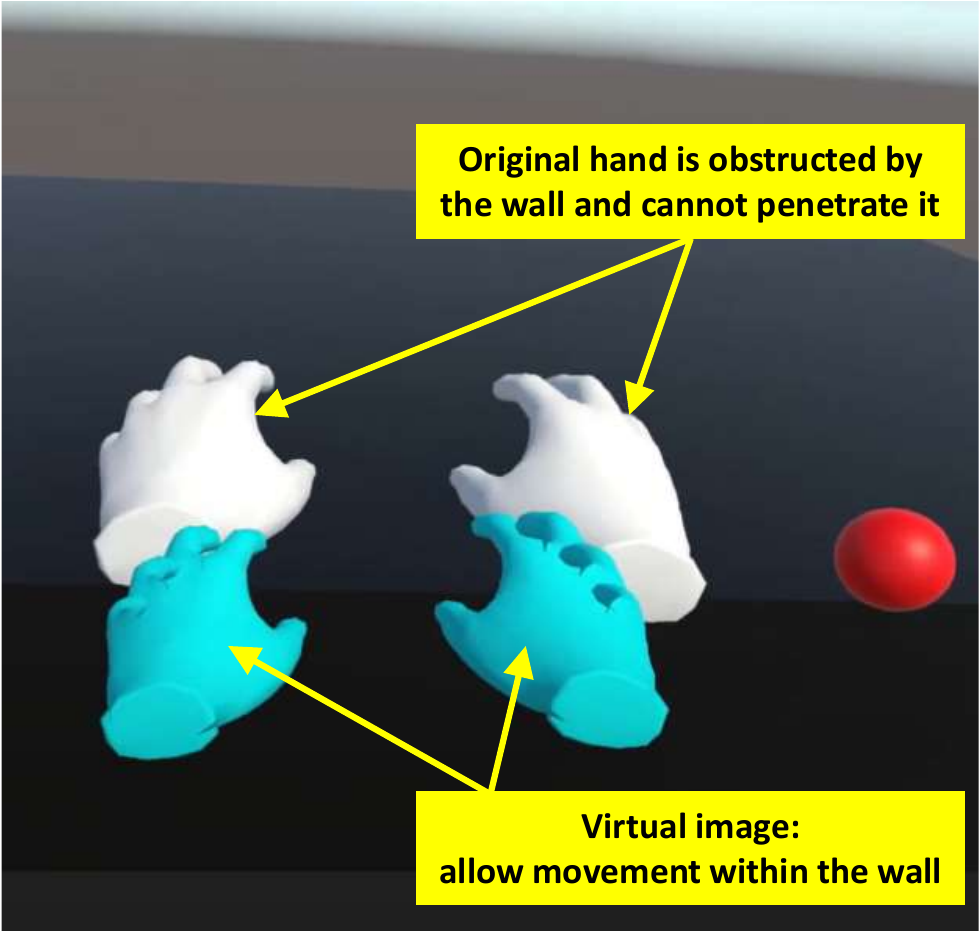}
  \caption{A solution for extending the user's hands into the walls}
  \label{fig:hands_wall}
\end{figure}

Additionally, users would inevitably extend their hands into the walls of the virtual space. If an attempt is made to illustrate this behavior by having the virtual hand collide with the wall, the absence of genuine haptic feedback will lead to an uncomfortable experience~\cite{Pruett2017}. To address this problem, the first common solution is to detect the intersection point between a user's hand and the virtual wall, gradually fading to black based on the penetration depth. As users push their hands further into the wall, the fading degree should increase. This design has an advantage in allowing users to learn that unintended intersections with the geometry of the virtual world result in fading, prompting them to self-correct their movements~\cite{Pruett2017}. Alternatively, another solution is to generate a virtual hand (represented in a different color) that allows movement within the wall, while the original hand is obstructed by the wall from penetration, as shown in Fig.~\ref{fig:hands_wall}.

\begin{figure*}[t]
  \centering
  \includegraphics[width=0.8\linewidth]{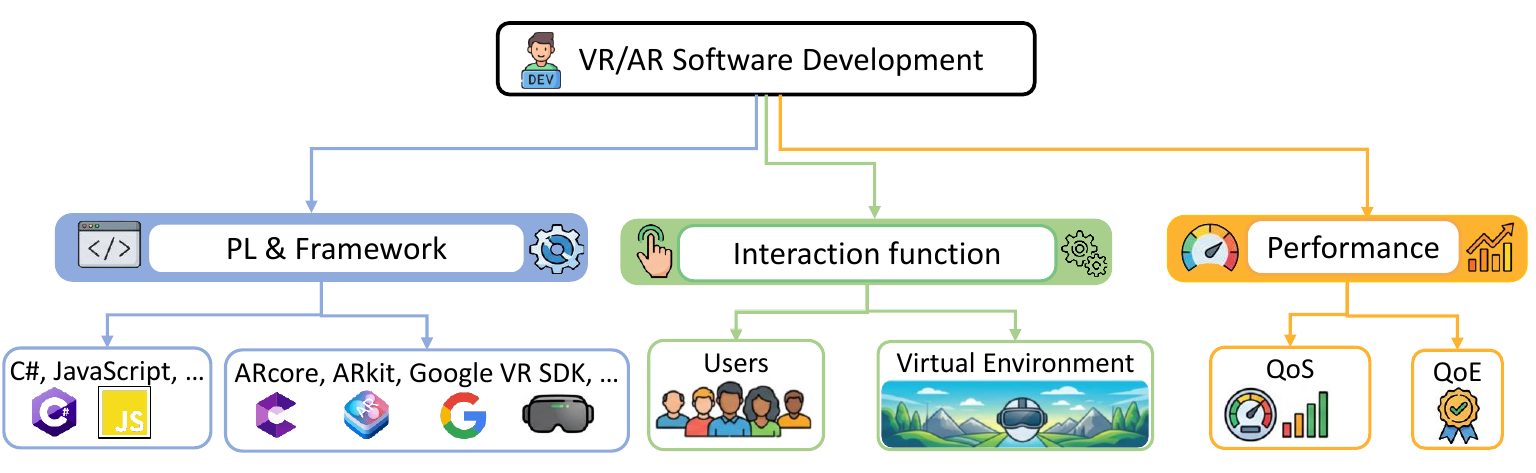}
  \caption{VR/AR software development}
  \label{Development}
\end{figure*}

\summary{\revise{Software development translates designs into functional VR/AR applications. Development environments are evolving from traditional 2D IDEs to immersive authoring 
3D tools. A key insight is the profound impact of LLMs and Generative AI, which enable in-situ prototyping and AI-driven speech interactions. Developers can now orchestrate 3D MR scenes directly through voice commands, fundamentally shifting the development paradigm toward intelligent, multimodal co-creation. Performance issues remain critical, focusing on both QoS metrics and QoE human factors.}}

\section{VR/AR Software Development}

In the phase of VR/AR software development, the research mainly consists of three parts: (1) issues caused by programming languages and frameworks in development; (2) the implementation of interactions; and (3) handling of performance issues on different platforms, as illustrated in Fig.~\ref{Development}.

\subsection{Development Frameworks \& Immersive Environments}
As mentioned in \cite{Naoures2018}, the most popular programming languages in VR/AR software development are C\# and JavaScript. As for the development framework, \cite{Trimananda2022} pointed out that Unity and Unreal Engine are the two most popular development frameworks. Meanwhile, some open-source libraries such as ARcore, ARkit, and Google VR SDK are adopted in VR/AR software development~\cite{Zainab2020, Sheena2022}. Some research has been proposed to analyze code from the languages and frameworks above. For example, \cite{Jacinto2021} proposed a code dependency analysis tool called VRDepend. VRDepend focuses specifically on direct associations, combinatorial relationships, and event-triggered relationships between VR objects (i.e., links between graphic units and script files), which are difficult to detect in traditional code dependency analysis tools.

Recently, the paradigm of VR/AR development is expanding from traditional 2D desktop environments to immersive authoring tools and VR-native Integrated Development Environments (IDEs). For instance, Giannaccari et al.~\cite{Mattia2025} proposed a VR-native IDE approach that leverages 3D visualizations (e.g., city metaphors) and spatial controller interactions to perform source code refactoring (e.g., moving classes and extracting methods) directly within a virtual environment. To lower the barrier for 3D content creation and accelerate rapid prototyping, several immersive authoring frameworks have also emerged. Shen et al.~\cite{Shen2025} introduced a VR authoring system tailored for 3D Gaussian Splatting (3DGS), enabling developers to perform fine-grained, point-level spatial editing and AI-driven generation. Similarly, to facilitate in-situ AR development, Kim et al.~\cite{Kim2025} and Lee et al.~\cite{Lee2025} integrated LLMs and generative AI to enable users to construct and edit 3D mixed reality scenes directly through voice commands. The former focuses on single-image MR reconstruction, while the latter addresses outdoor scene understanding and multi-agent AI orchestration. Additionally, Luo et al.~\cite{Luo2025} developed an immersive authoring tool specifically for prototyping ``remapped interactions'' (e.g., redirection, resizing), helping developers intuitively visualize and mitigate sensory conflicts during the development phase.

\begin{figure*}[h]
  \centering
  \includegraphics[width=0.8\linewidth]{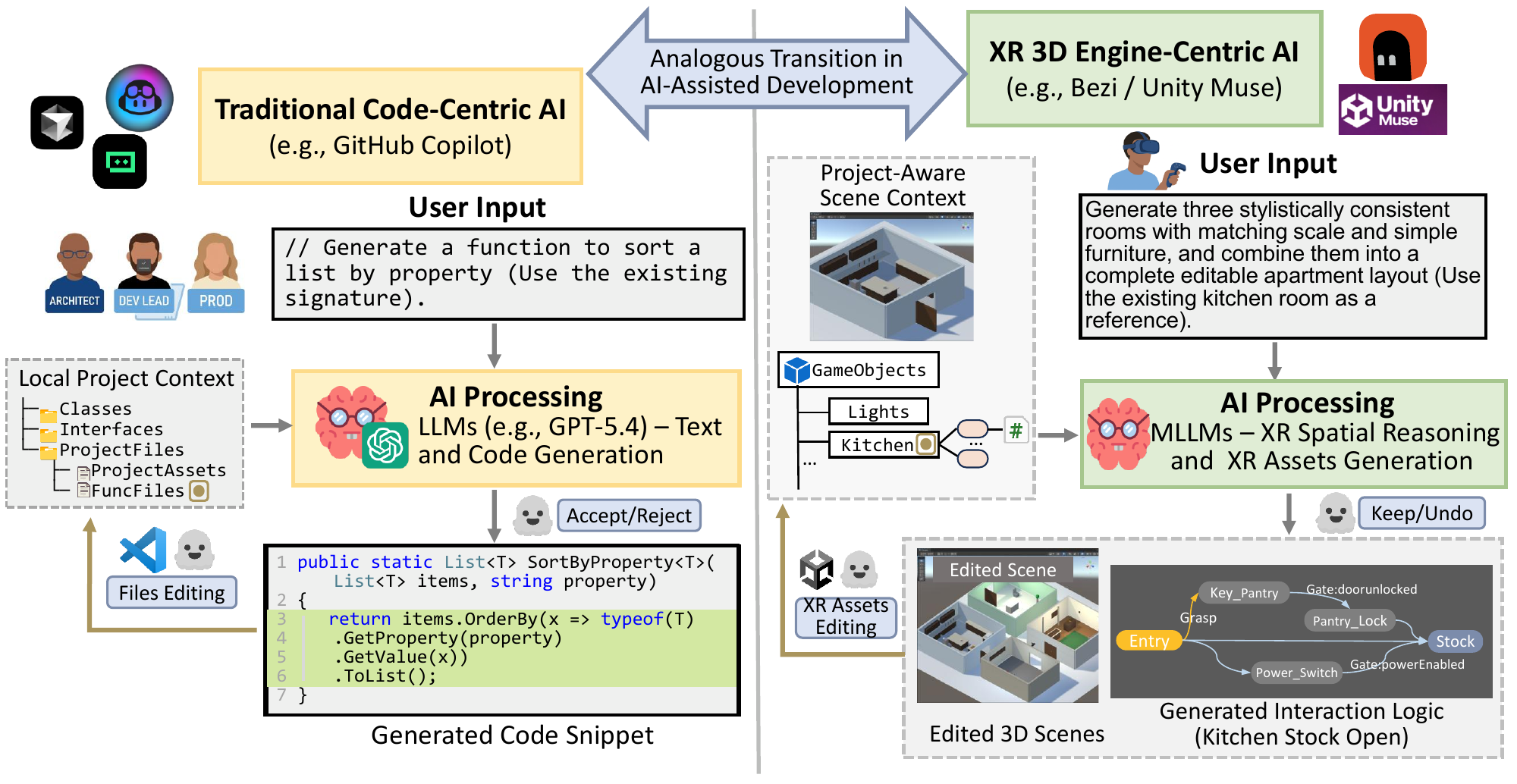}
  \caption{From code-centric AI assistants to XR 3D engine-centric AI: An analogous transition toward project-aware spatial reasoning, asset generation, and interaction logic synthesis}
  \label{fig:bezi_zhengyang}
\end{figure*}

Beyond research prototypes for immersive authoring, a parallel industrial trend is embedding AI assistance directly into mainstream engine workflows. As illustrated in Fig.~\ref{fig:bezi_zhengyang}, this trend can be understood as an analogous transition from traditional code-centric AI assistance—primarily focused on local code context, text prediction, and source-file editing—to XR 3D engine-centric AI, where assistance becomes project-aware and operates over scenes, assets, object hierarchies, interaction logic, and engine-native configurations. Recent tools, such as Bezi for Unity~\cite{Bezi2026}, Unity Muse/Unity AI~\cite{UnityMuse2023, UnityAI2026}, and Epic Developer Assistant for Unreal Engine~\cite{EpicAssistant2025, EpicStateUnreal2025, EpicForum2025} suggest that AI support in XR development is shifting from \textit{isolated asset generation} toward \textit{integrated assistance} for debugging, code generation, asset creation, scene editing, and rapid prototyping within the engine environment. This shift is particularly important for XR development, where implementation is inherently cross-modal and tightly coupled across scripts, scene objects, assets, interaction logic, and engine-specific settings. In a similar vein, editor-integrated asset-generation tools, such as the Scenario Unity Plugin~\cite{ScenarioUnity2025}, further lower the barrier to iterative content production within existing development pipelines. Taken together, these developments suggest that VR/AR development environments are increasingly evolving into AI-mediated authoring ecosystems, in which generative AI models function less as standalone content generators and more as workflow copilots for project-aware, end-to-end implementation and iteration.

\subsection{Implementation of Interaction}

The implementation of interaction refers to the interaction among multi-users and the interaction between users and the virtual scene. As for the multi-user interaction, network configuration is an important development module since it affects the quality of multi-user interaction. Some research has focused on this topic. For example, \cite{Cheng2022} conducted a survey of social VR platforms and proposed an in-depth analysis of Meta's VR application called Horizon Workrooms, revealing its network protocol usage and requirements through measurement studies. Specifically, the study exposed the network protocols used by Workrooms, including UDP streams and WebRTC streams, and understood how Workrooms handles virtual content delivery, real-time multimedia content delivery, and how it affects bandwidth requirements as users increase. Cheng et al.~\cite{Cheng2022_2} conducted a survey of metaverse and proposed a reality check by comparing the network operation and performance of two representative platforms, Meta's Horizon Workrooms and Microsoft's AltspaceVR, which in turn discussed the technological challenges, opportunities, and future research directions for the metaverse. As for the interaction between users and the virtual scene, some privacy issues have been focused on. Lehman et al.~\cite{Lehman2022} developed a framework to avoid `hidden operation' threats. The details are shown in \S~\ref{external}.
Moreover, implementing natural and seamless interactions between users and complex virtual/physical environments has driven recent innovations. On one hand, physical-to-virtual spatial mapping interactions have been systematically supported by tools like RemapVR~\cite{Luo2025}, which formalizes the implementation of repositioning and redirection interactions to overcome physical space limitations. On the other hand, AI-driven multimodal interactions are becoming a new standard. Systems such as ImaginateAR~\cite{Lee2025} and RealityCrafter~\cite{Kim2025} implement speech-driven interaction pipelines using LLMs, allowing developers and users to naturally dictate commands to generate, manipulate, and remove 3D assets in real-world AR scenes, fundamentally changing how interaction logic is implemented and executed.

\subsection{Performance Issues}
\begin{figure}[h]
  \centering
  \includegraphics[width=\linewidth]{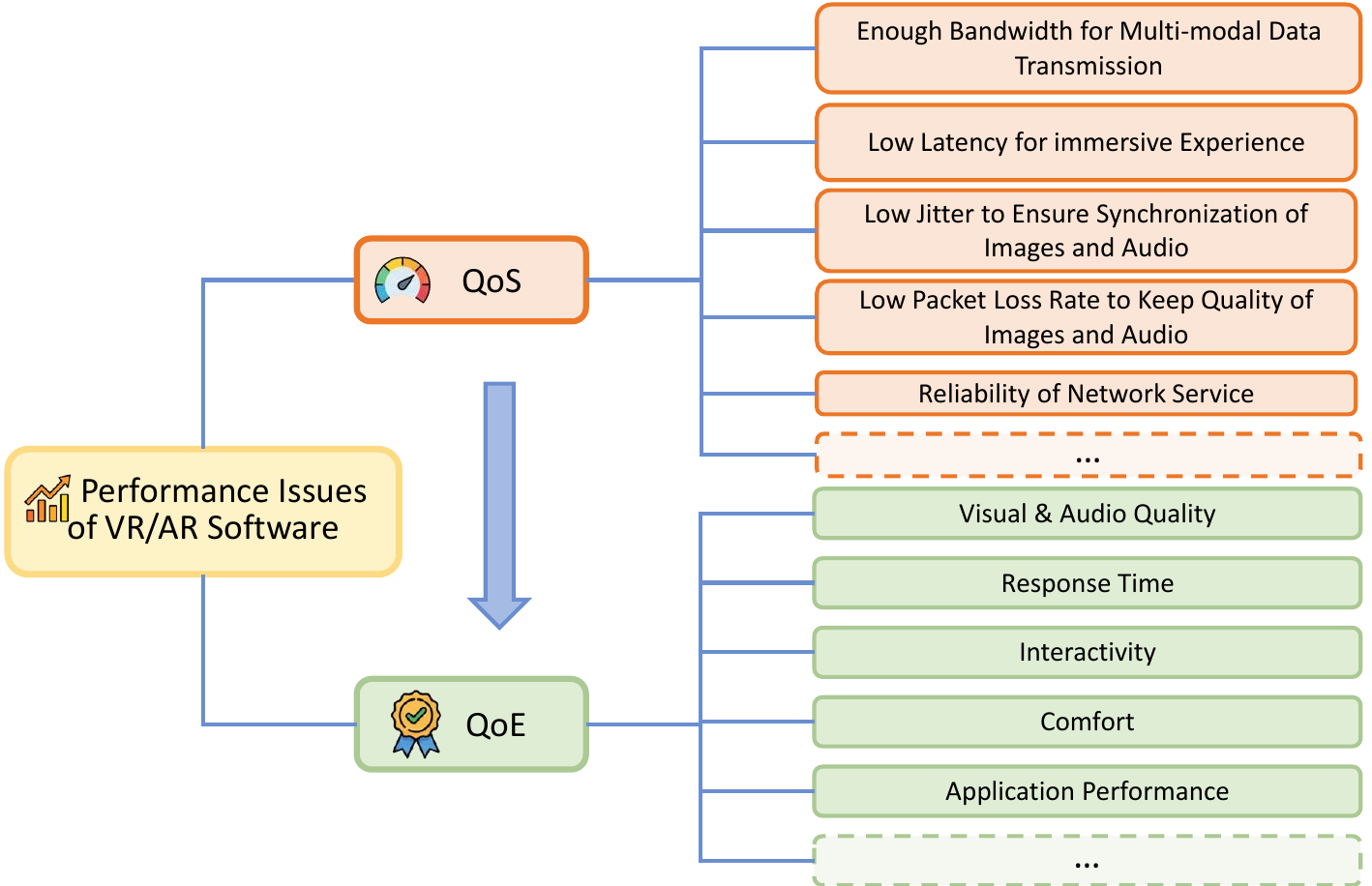}
  \caption{Categorization of VR/AR performance issues }
  \label{performance}
\end{figure}

Nowadays, much research focuses on addressing performance issues in the development phase since it is important to evaluate the quality of VR/AR software. Due to immersive user experience, the VR/AR software quality can be measured by quality of service (QoS) \cite{Van2024} and QoE~\cite{Debarba2024}. Correspondingly, the performance issues of VR/AR systems typically involve these two metrics. Fig.~\ref{performance} depicts the categorization of these performance issues related to VR/AR software systems. 

\textbf{QoS} is the quality of service that a network or service provider is able to provide when transmitting data. It is an objective metric that is usually used to assess and guarantee service performance through technical indicators. In the VR/AR software, QoS involves several technical parameters, including but not limited to:
\begin{itemize}[leftmargin=*, labelindent=0pt,noitemsep,topsep=0pt]
\item \textit{Bandwidth}: VR and AR applications typically require the transmission of large amounts of data, including high-resolution images, video, and audio, which requires a network with sufficient bandwidth to support high data rates.
\item \textit{Latency}: Latency is one of the most critical QoS metrics in VR and AR applications. To provide an immersive experience, the system needs to respond to user actions and inputs with very low latency. Typically, latency needs to be less than 20 milliseconds to prevent users from feeling dizzy or uncomfortable.
\item \textit{Jitter}: Jitter refers to the variation in packet arrival times. In VR and AR applications, jitter can cause images and audio to be out of sync, affecting the user experience. Therefore, the network needs to keep jitter low.
\item \textit{Packet Loss Rate}: Packet loss can lead to degradation or interruption of image and audio quality. In VR and AR applications, packet loss rates need to be kept very low to ensure a smooth experience.
\item \textit{Reliability}: The stability and availability of network services is critical for VR and AR applications. Any disruption can significantly affect the user experience.
\end{itemize}

\textbf{QoE} is the overall experience and satisfaction of a user when using a VR/AR software. It is a subjective metric based primarily on user perceptions and expectations. QoE in the VR/AR software involves a number of factors, including but not limited to:
\begin{itemize}[leftmargin=*, labelindent=0pt,noitemsep,topsep=0pt]
\item \textit{Visual Quality}: High resolution and high frame rate images are key to delivering an immersive experience. Users expect clear, smooth and distortion-free images. Frame rates of 90 frames per second (FPS) or higher are typically required to avoid visual lag and discomfort.
\item Audio Quality: High-quality 3D audio is an important factor in enhancing immersion. Users expect clear, synchronized, and spatial sound.
\item \textit{Response Time}: The latency between the user's actions and the system's response needs to be as low as possible. Any noticeable delay will break immersion and cause discomfort for the user.
\item \textit{Interactivity}: Users expect a natural and seamless interaction with the virtual environment. This includes the accuracy and immediacy of gesture recognition, head tracking, and other input methods.
\item \textit{Comfort:} Comfort is an important QoE metric in VR and AR applications. Prolonged use should not lead to discomfort or fatigue for the user. This is closely related to factors such as visual quality, response time and device design.
\item \textit{Other Application Performance:} Other application performance involves Application startup time, load time, and smoothness of operation also affects the user experience. Users expect fast startup and seamless operation.
\end{itemize}

Existing research focuses on performance issues based on the above two perspectives.
In particular, \cite{Nusrat2020, 9402052} conducted an empirical study about performance issues in VR/AR Unity projects. After collecting a dataset of 230 performance bug fix commits from 100 open-source Unity projects on GitHub, they summarized the Unity performance bug category through manual analysis. Based on the AST-level code change analysis, they found that Unity's callback methods (e.g., Update, Start, etc.) are susceptible to performance problems and that Unity performance fix commits are high due to code dependency issues. Additionally, through code change analysis they found that performance fixes can be a combination of source code and prefabricated files, or sometimes just prefabricated files. These findings are useful for developers and future researchers. Moreover, \cite{Antonio2020} proposed a static analysis tool called UnityLinter, designed to help Unity video game developers detect seven ``Bad Smells'', which are relevant to video game development. These Bad Smells were defined by analyzing existing literature and discussion forums, and their relevance was verified through surveys. The researchers then analyzed the occurrence of these Bad Smells in 100 open-source Unity projects and evaluated the accuracy of UnityLinter. The seven Bad Smells include some performance-related issues. Further, \cite{hogan2022} proposed an empirical study aimed at analyzing the performance of Unreal Engine (UE)-based VR applications. The researchers collected performance optimization changes from seven UE XR projects, 78 UE XR discussion questions, and three UE documentation sources. From the collected dataset they identified 14 performance error types, and developed a static analyzer, UEPerfAnalyzer, for detecting performance errors in configuration files and source code.

\summary{\revise{Software development is the specific process that follows requirements analysis and software design. Existing VR/AR research mainly focuses on development frameworks, interaction implementation, and performance issues. Recently, development environments have evolved from traditional 2D IDEs to VR-native IDEs and immersive authoring tools (e.g., integrating LLMs and generative AI for in-situ prototyping). Interaction implementation focuses on multi-user networking, physical-virtual spatial remapping, and AI-driven speech interactions. Performance plays a crucial role in VR/AR software development, and it is related to QoS-related metrics and QoE-related human factors. These advancements collectively help developers achieve rapid, efficient, and immersive software development.}}

\section{Testing of VR/AR}
In this section, we first introduce bugs in VR/AR software development. Then, we will illustrate different testing methods applied in VR/AR. 

\subsection{Bugs \& Test Smells in VR/AR Applications}
There already exist empirical studies on addressing the common issues in open-source VR/AR applications in the wild. Due to the high computational cost of real-time rendering and the resource limitations of wearable devices, performance overheads are widely recognized throughout the process of VR/AR software engineering practice. In~\cite{9402052}, 183 real-world performance optimizations from 45 VR software projects are analyzed and categorized into 11 categories. Then, the authors apply static analysis to determine how these optimizations affect different life-cycle phases of VR applications. Among all of them, graphics simplifications, rendering optimizations, and heap avoidance are three dominant ones, which account for more than half of the studied optimizations. Moreover, these three categories differ from conventional software. 

\begin{figure}[t]
  \centering
  \includegraphics[width=\linewidth]{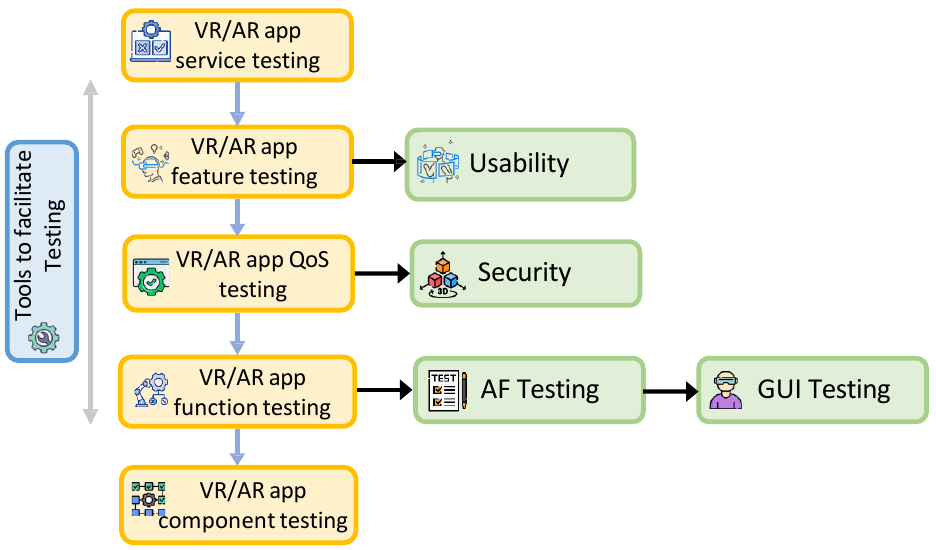}
\caption{Testing procedure in VR/AR applications.}
  \label{test}
\end{figure}

Generally, VR/AR applications require specific hardware devices and various platforms. For instance, VR/AR applications can be downloaded from certain app stores, such as the  Meta Horizon Store, and used on VR/AR devices like Meta Quest 2 and Meta Quest 3. However, this may discourage users from accessing VR/AR applications. For this reason, \textit{the WebXR Device API}~\cite{webxr} was released to enable XR applications to be deployed on the web. However, a large number of WebXR applications suffer from various bugs that can deteriorate the user experience or cause severe issues. In \cite{9251053}, the authors conduct empirical studies of WebXR bugs by collecting 368 bugs from 33 WebXR projects on GitHub and analyzing the issue reports and release notes. First, they build a bug taxonomy based on the symptoms and root causes of the collected bugs. Then, they compare these WebXR bugs with traditional JavaScript programs and web applications to observe their differences. One prominent difference for WebXR is the existence of incompatible devices and improper handling of diversified user interactions. Such bugs are never addressed in traditional JavaScript programs or web applications. As a result, a large number of severe bugs and issues in XR applications never occur in traditional software applications due to the uniqueness of XR applications.

Beyond traditional software bugs, the quality of VR test code itself has garnered attention. Gurramkonda et al.~\cite{Faraz2025} investigated test smells in Unity-based VR projects, revealing that developers often write fewer tests with suboptimal designs. They proposed VRTestSniffer, a static analysis tool leveraging Abstract Syntax Trees (AST), Control Flow Graphs (CFG), and Data Flow Graphs (DFG) to detect 17 VR-specific test smells. Their empirical analysis further highlighted that certain test smells are closely associated with production code design issues, such as Blob and Complex Classes, emphasizing the need for dedicated test quality assurance in VR.

\subsection{Testing in VR/AR Applications}
Due to the different characteristics of VR/AR techniques, new testing methods are needed to ensure the quality of software engineering. Fig.~\ref{test} summarizes the overall procedure and our concentrated issues in VR/AR app testing. These aspects are secured by the following testing methods: Application-specific Functional (AF) testing, GUI testing, Usability testing (user experience), and Security testing. We will also include different tools that facilitate testing in this domain.

\subsubsection{Application-specific Functional (AF) Testing}
\begin{figure*}[!t]
\centering
\subfloat[Event tests from MixedRealityToolkit-Unity~\cite{unity}]{\includegraphics[width=2.5in]{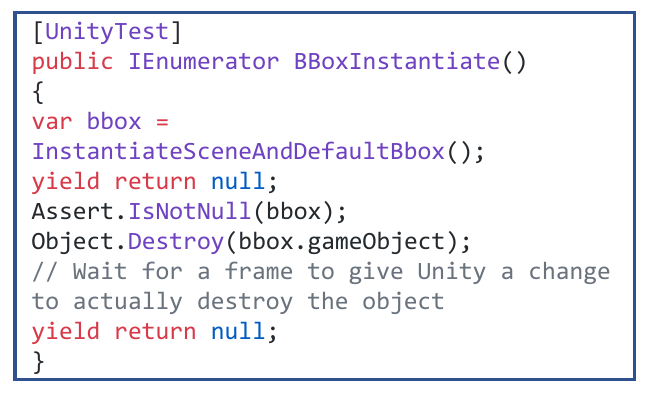}%
\label{code1}}
\hfil
\subfloat[GUI tests from MixedRealityToolkit-Unity~\cite{unity}]{\includegraphics[width=3in]{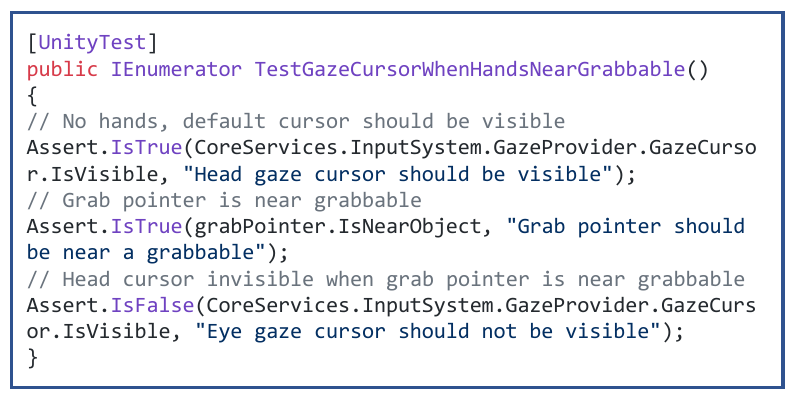}%
\label{code2}}
\caption{Examples of test xases}
\label{fig:example}
\end{figure*}

The empirical study~\cite{rzig2023virtual} of automated testing on Unity-based VR applications has analyzed all automatic tests among 314 open-source VR applications. They state that 79\% of the VR projects examined lack automatic tests, and for the rest of these projects, the median ratios of functional-methods to test-methods are much lower than those of other project types. For these AF issues, app logic tests are categorized based on specific logic and functionalities of VR applications. After scrutinizing these tests, they further divide these into seven sub-categories: Event Tests, Authentication Tests, Exception Tests, Configuration Tests, I/O Tests, Location Tests, and others. Fig.~\ref{fig:example}\subref{code1} presents an event test from an open-source MR application. It verifies if we can instantiate a bounding box at runtime. This taxonomy establishes the foundations of VR testing practices on AF issues. 
Souza \etal~\cite{correa2018automated} proposed an approach to automate the generation of test requirements and test data using a semiformal language. In their design, they first utilize requirements specification and domain knowledge to illustrate the requirements of VR applications. Then, they map the requirements using a semiformal language called BeLaRS into a requirement flow graph (RFG). Next, test data is generated to fulfill the test requirements based on the RFG given the test criterion. Two experiments are conducted to validate the approach. The results show the usability and effectiveness of this approach. 

Furthermore, to validate the unique spatial interactions in VR (e.g., 6DoF controller gestures), Gu and Rojas~\cite{Gu2025} introduced INTENXION, a test automation framework for Unity XR applications. Based on a comprehensive taxonomy of XR user interactions, INTENXION provides specific APIs to simulate complex compositional actions (e.g., grab, trigger, rotate) and assertions to automatically validate expected outcomes, thereby bridging the gap between traditional 2D UI testing and immersive spatial VR interaction testing.

\subsubsection{GUI Testing} Testing methods of GUI can mainly be divided into two categories: static and dynamic testing. For static testing, in the work of~\cite{rzig2023virtual}, they have emphasized that GUI tests are one of the VR-attention categories compared with traditional software. GUI tests are generated to evaluate the layouts, input fields, and text of the application's interface. Fig.~\ref{fig:example}\subref{code2} presents a GUI test. It tests the status of the cursor under different circumstances. To categorize the test type, the researchers manually analyze 370 test methods from 36 VR projects. However, the number of GUI tests found is only 8. In this sense, it is easy to conclude that VR software engineers pay little attention to static testing on the GUI. 

Researchers have developed methods for automated dynamic testing for VR scene exploration. For example, Wang \etal~\cite{wang2022vrtest} proposed a testing framework called VRTest to automatically control the player's camera and interact with the environment. It uses two strategies: VRMonkey and VRGreed. In experiments, both strategies successfully identified interactable objects. However, the VRGreed strategy proved to be more effective. In 2023, Wang \etal ~\cite{wang2023vrguide} further improved the efficiency of their framework. They state that VRTest fails to address VR-specific challenges, such as avoiding object occlusion and movement. As a result, they present VRGuide, which utilizes \textit{Cut Extension} as their technique to guide their testing camera into an optimized route. They compare VRGuide and VRTest based on eight VR software projects. They use object coverage, method coverage, and the ability to detect bugs to evaluate their performance. The results show that VRGuide is more efficient than VRTest in terms of the object coverage and method coverage in most of the VR projects. What's more, VRGuide is able to discover real-world issues that have been confirmed and fixed by developers. 


To further tackle the complexity of 3D virtual environments and diverse interaction types, Zhu et al.~\cite{Zhu2025} proposed VRExplorer. It utilizes an Entity-Action-Task (EAT) framework to model VR interactions generically and employs a Probabilistic Finite State Machine (PFSM) alongside Unity's NavMesh for autonomous scene exploration and decision-making. Evaluations demonstrated that VRExplorer significantly outperforms VRGuide in both executable code and method coverage, and it successfully uncovered deep functional bugs in real-world VR projects. 


\subsubsection{Usability Testing (User Experience)} \label{sec:UsabilityTesting} 
\begin{figure*}[!t]
\centering
\subfloat[Wrong virtual object placement in AR apps~\cite{rafi2022predart}]{\includegraphics[width=2.95in]{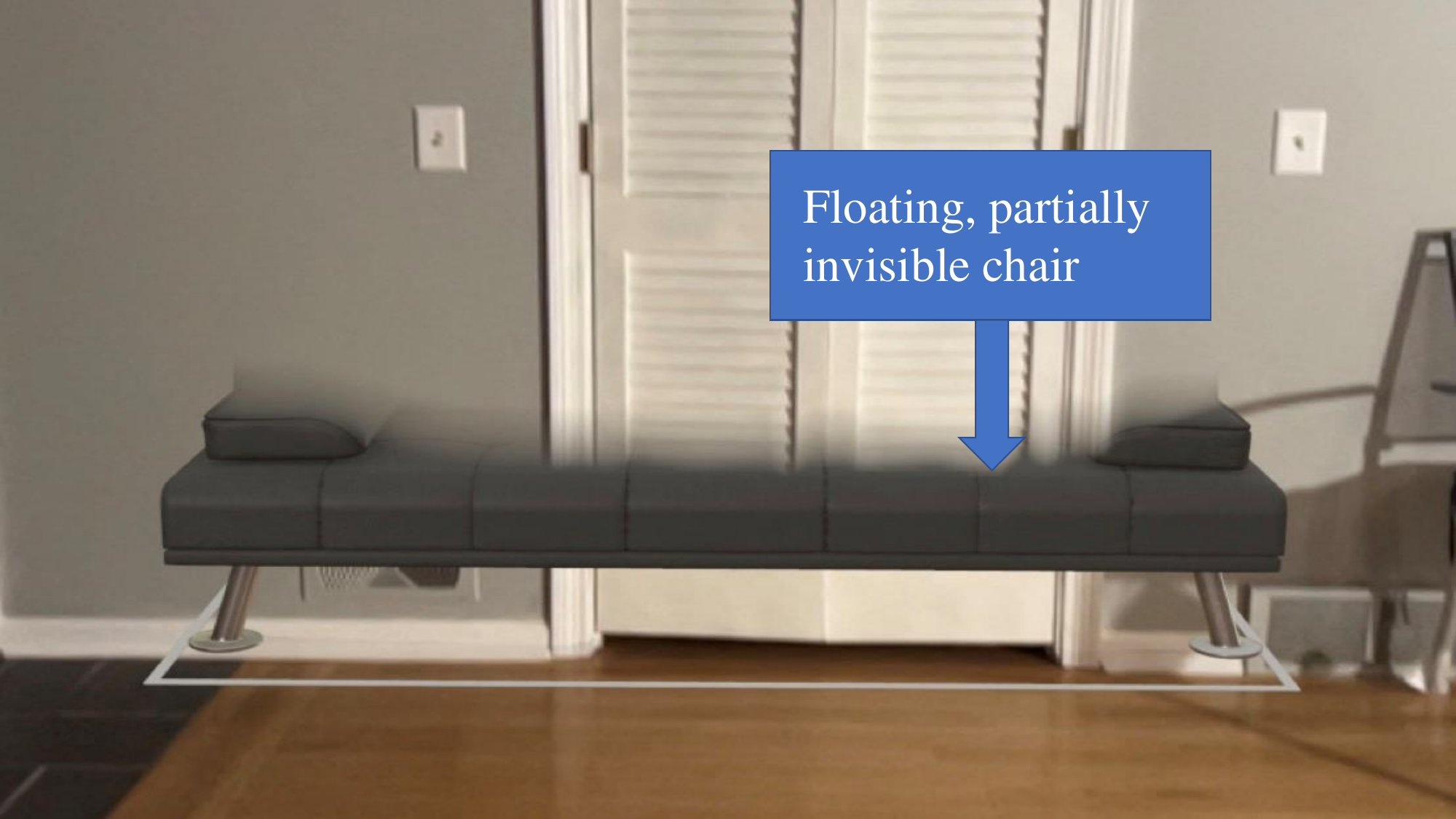}%
\label{issue2}}
\hfil
\subfloat[A bug causing SVI issues in SteamVR app~\cite{issue}.]{\includegraphics[width=3in]{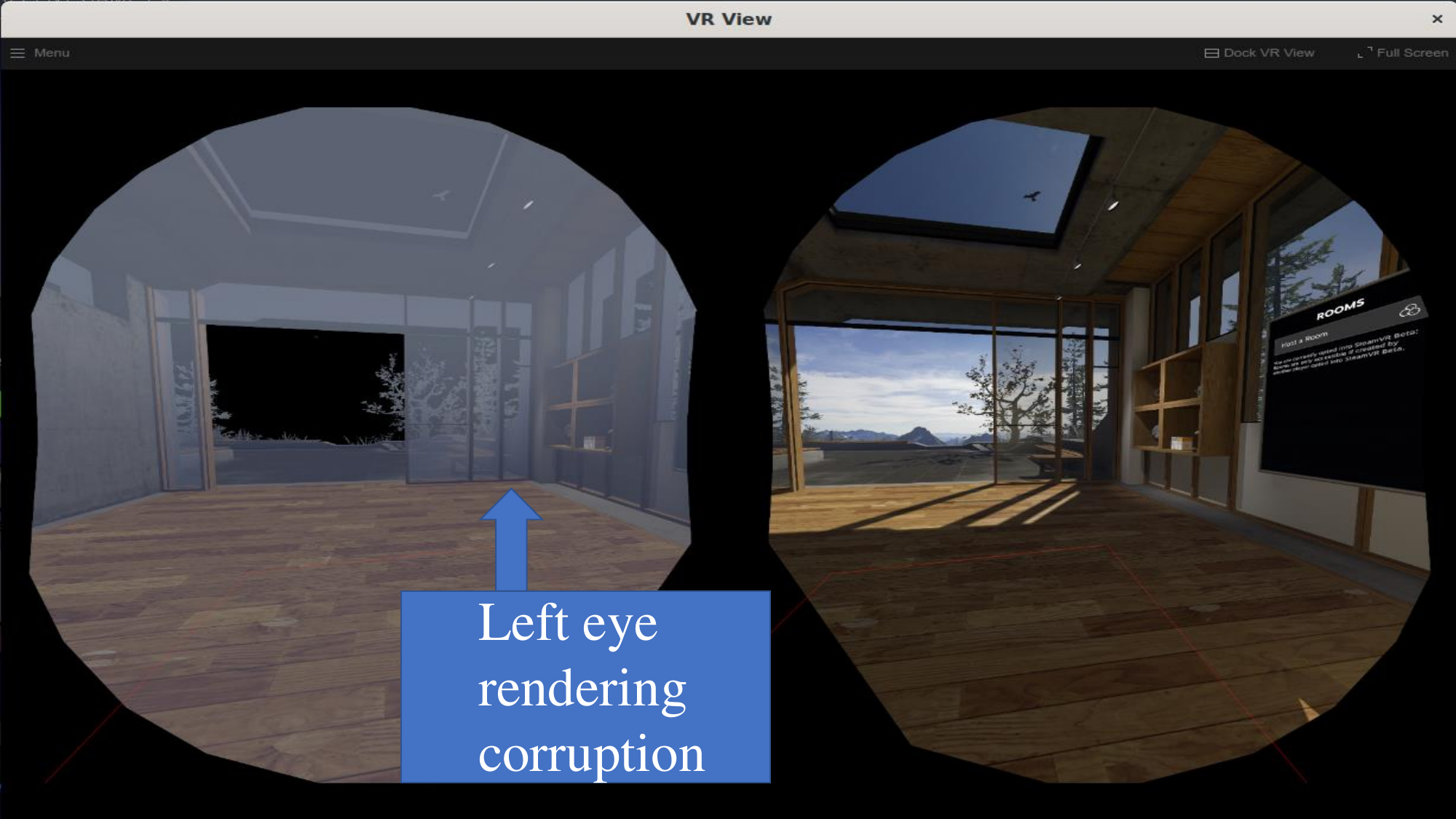}%
\label{issue1}}
\caption{Examples of visual issues}
\label{fig:visual}
\end{figure*}

Though visual functions can be validated through GUI testing, there exist issues in virtual objects and scenes that may be harmful to user experience. The first is the wrong virtual object placement issue. Fig.~\ref{fig:visual}\subref{issue1} shows us a typical example that can occur in real-world AR apps. To evaluate the quality of virtual object placement, Rafi \etal ~\cite{rafi2022predart} developed an approach called PredART to predict human ratings of virtual object placements in 2022. They first take screenshots using an automated moving camera of different virtual objects' placement status. Then, they use another platform to help them label these pictures. Next, they train a hybrid deep neural network with CNN and MLP for further prediction. Rafi \etal introduce different metrics to assess the overall prediction accuracy and compare it with a state-of-the-art image regression approach called ResNet. Also, they apply PredART to detect bad object placements with real-world VR apps. It is proven to be effective on their testing data and superior to ResNet. Moreover, PredART can effectively identify non-realistic screenshots in real-world VR apps. In 2024, Yang \etal ~\cite{yang2024towards} propose a novel approach called VOPA to further improve the PredART to expand it into real-world AR scenes. They extract runtime metadata and design different mutations of imprecise object placements using APIs in Google ARCore based on AR apps. Then, they label the matched screenshots and metadata following the method Rafi \etal used. As for the classification model, they design a multimodal model to utilize both screenshots and metadata as inputs. As what Yang \etal ~\cite{yang2024towards} state, VOPA can do better than PredART to tackle the randomness and noises in real-world AR scenes. 

Another severe issue that will degrade the quality of GUI rendering is called stereoscopic visual inconsistency (SVI). This phenomenon, which is caused by false rendering for user's left and right eyes will cause discomfort and even bring bad health effects to users. Fig.~\ref{fig:visual}\subref{issue2} presents one example of an SVI issue. To state this problem, Li \etal ~\cite{li2024less} conduct an empirical study on SVI bug reports collected from collected VR platforms. They first collect 15 real-world VR platforms from VR forums and app stores. Then, they utilize keyword search to filter out potential bug reports of SVI issues. After that, they summarize the collected SVI issues into two categories: view-level and object-level inconsistencies. To detect such issues, they developed an unsupervised testing framework called StereoID. Their framework first learns the relations between screenshots of two eyes and tries to generate the right-eye images given the left-eye images. A distance calculator is designed to compute the distance between the generated right-eye image and the real right-eye image. Last, the distance metric will be used to identify the inconsistency. Li \etal perform several experiments to evaluate the ability of their framework to generate stereo images and detect SVI issues. The results validate the effectiveness and usefulness of StereoID based on the datasets they collected. 

  

To combine software testing goals and user experience, Rui \etal \cite{prada2020agent} create an autonomous cognitive test agent that can both explore the interactions given a specific test goal and assess the user experience while testing. They introduce \textit{Belief-Desire-Intent} (BDI) model to set up goals and tactics for their test agent to interact with the VR environment. To perform user experience tests, FAtiMA~\cite{mascarenhas2018virtual}, which implements the OCC model has been utilized as an extension in BDI platforms. During their interactions with the environments with testing goals, the agents will also check the status of their internal state of emotions, which have built a solid foundations for further user experience evaluation in AR/VR testing. Another testing framework called ARCHIE~\cite{lehman2020archie} collects user feedback and system state data during testing to identify important issues during run-time. It supports performance and usability feedback from users through a smart-watch based app. ARCHIE will aggregate run-time system data with poor performance or usability feedback submitted by users for further analysis. Also, it can compare different function options from multiple version for each application at run-time. The authors evaluate ARCHIE based on different types of AR applications. The results show that it is efficient for ARCHIE to provide evidence to identify and debug user experience (UX) issues. Also, there is no significant overhead to use ARCHIE in run-time AR/VR apps.

\subsubsection{Security Testing}
3D models are important assets in AR/VR engineering. However, they can be easily extracted, duplicated, or cloned, posing a security threat. Zuo \etal ~\cite{zuo2023peek} conduct the first systematic study aiming at 3D model cloning in 3D gaming. They propose 3DScan to detect such phenomenon in 3D mobile games based on Unity. It first extract 3D models using AssetStudio, given the APK file during compilation. Then, they design a normalization-based index algorithm to detect 3D model clones based on raw data from AssetStudio. Lastly, 3DScan identifies 3D model cloning using different methods for static and animated models with tested 176,361 Unity-based games. The results suggest that 63.03\% static 3D models and 37.07\% animated 3D models are potential subjects for cloning. They also do experiments to validate the accuracy of 3Dscan. Their study reveals the presence of unauthorized copies despite common 3D models. 

\subsubsection{Tools to Facilitate Testing}
There are research works that aim to provide support for VR testing in different processes. Qin \etal ~\cite{qin2022dytrec}
proposed a dynamic testing recommendation tool by analyzing official APIs from the official documentation. They first collected all Unity APIs that may contain dynamic updating features and then categorized them into different types. Then, they analyzed the source code from target VR projects based on the categorization and recommended test types. They applied their tools to 20 VR projects. The results have recommended 39 testing suggestions with dynamic testing types such as animation, collision, and general, which are linked to target methods or classes. Another framework called ORIENTER~\cite{li2024groundedguiunderstandingvision} is proposed to detect interactable GUI elements (IGEs) on 3D rendering. They incorporate context information in VR apps with zero-shot large multimodal models (LMMs) to facilitate their tasks of IGE detection. They conducted extensive experiments to validate the effectiveness of ORIENTER, which surpasses state-of-the-art GUI element detection approaches. The tool can also be regarded as a cornerstone for further research on automatic testing and GUI search.

Similarly, the rapid advancement of LLMs has inspired new paradigms for VR exploration testing. Qin et al.~\cite{Qin2024, qi2026harnessing} conducted extensive case studies on utilizing generative AI (e.g., GPT-4o) for Field of View (FOV) analysis in black-box VR testing. They demonstrated that through structured chain-of-thought prompt engineering, LLMs can effectively identify testing entities, extract core visual features (e.g., color, placement, shape), and understand 3D spatial relationships directly from 2D screenshots without requiring the underlying source code. Although challenges remain in precise coordinate localization and reasoning reliability, these studies highlight the immense potential of multimodal LLMs acting as autonomous agents for dynamic VR GUI exploration.

\summary{\revise{According to empirical studies on common issues in AR/VR applications, severe bugs and test smells are quite different from those in traditional software. To assure the quality of AR/VR applications, testing methods are divided into AF testing, GUI testing, usability testing, and security testing. Recent advancements have evolved from basic script examinations to sophisticated spatial interaction automation (e.g., testing 6DoF controller inputs) and model-based autonomous scene exploration. Furthermore, to overcome the limitations of white-box testing, researchers are increasingly harnessing LLMs and zero-shot LMMs to facilitate dynamic testing, GUI element detection, and FOV-based environment understanding, pushing VR testing towards intelligent, automated black-box paradigms.}}

\section{Deployment of VR/AR Software} 
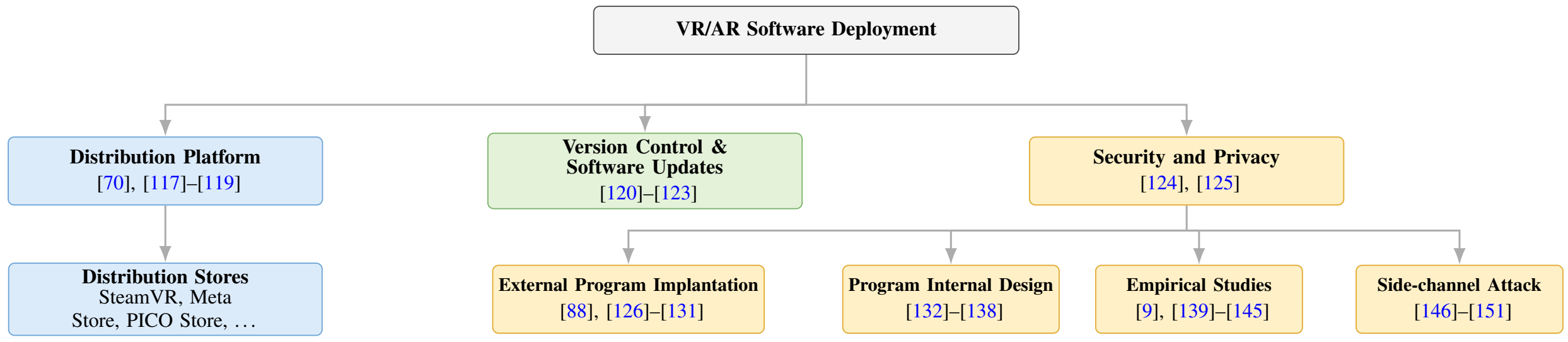
\begin{figure*}[t]
  \centering
  \resizebox{0.90\textwidth}{!}{%
  \begin{tikzpicture}[
    font=\scriptsize,
    conn/.style={draw=gray!65, line width=0.65pt, -{Latex[length=2mm,width=1.5mm]}},
    branch/.style={draw=gray!65, line width=0.65pt},
    root/.style={draw=black!70, fill=deploygray, rounded corners=2pt, align=center,
      text width=5.0cm, minimum height=0.58cm, inner sep=2pt},
    topic/.style 2 args={draw=#1, fill=#2, rounded corners=2pt, align=center,
      minimum width=3.8cm, minimum height=0.82cm, inner sep=2pt},
    leaf/.style 2 args={draw=#1, fill=#2, rounded corners=2pt, align=center,
      text width=3.65cm, minimum height=0.82cm, inner sep=2pt},
    securityleaf/.style={draw=deployorangeframe, fill=deployorange, rounded corners=2pt,
      align=center, minimum width=2.5cm, minimum height=0.82cm, inner sep=2pt}
  ]
    \node[root] (root) at (1.95,0) {\textbf{VR/AR Software Deployment}};

    \node[topic={deployblueframe}{deployblue}] (platform) at (-5.8,-1.70)
      {\textbf{Distribution Platform}\\[1pt]
      \mbox{\cite{Trimananda2022,zhan2024,Yang2023,Meng2026}}};
    \node[topic={deploygreenframe}{deploygreen}] (version) at (0,-1.70)
      {\textbf{Version Control \&}\\[-1pt]\textbf{Software Updates}\\[1pt]
      \mbox{\cite{Unity2023,Meta2023,Simhadri2025,wu2026xrfix}}};
    \node[topic={deployorangeframe}{deployorange}] (security) at (6.55,-1.70)
      {\textbf{Security and Privacy}\\[1pt]
      \mbox{\cite{teymourian2025sok,Dennis2025}}};

    \node[leaf={deployblueframe}{deployblue}] (stores) at (-5.8,-3.25)
      {\textbf{Distribution Stores}\\[-1pt]
      SteamVR, Meta Store, PICO Store, \ldots};

    \node[securityleaf] (external) at (-0.20,-3.25)
      {\scalebox{0.92}{\textbf{External Program Implantation}}\\[1pt]
      \mbox{\cite{Lee2021,Zhang2023,yang2024,Lehman2022,Cheng2024,Mengascini2024,mukherjee2025shadowed}}};
    \node[securityleaf] (internal) at (3.70,-3.25)
      {\scalebox{0.92}{\textbf{Program Internal Design}}\\[1pt]
      \mbox{\cite{Carter2024,Nair2023,Samaikya2022,Kim2023,Ruth2019,Zuo2022,Pan2025}}};
    \node[securityleaf] (empirical) at (6.70,-3.25)
      {\scalebox{0.92}{\textbf{Empirical Studies}}\\[1pt]
      \mbox{\cite{Ananya2018,noah2022,Pınar2022,Guo2024,Guo2025,Cai2020,kim2025autovr,shoaib2025principled}}};
    \node[securityleaf] (sidechannel) at (9.85,-3.25)
      {\scalebox{0.92}{\textbf{Side-channel Attack}}\\[1pt]
      \mbox{\cite{Meteriz2022,Luo2022,luo2020,Falk2021,Nahuel2021,Arafat2021}}};
    \coordinate (rootBranch) at (1.95,-0.90);
    \draw[branch] (root.south) -- (rootBranch);
    \draw[conn] (rootBranch) -| (platform.north);
    \draw[conn] (rootBranch) -| (version.north);
    \draw[conn] (rootBranch) -| (security.north);

    \draw[conn] (platform.south) -- (stores.north);

    \coordinate (securityBranch) at (6.55,-2.42);
    \draw[branch] (security.south) -- (securityBranch);
    \draw[conn] (securityBranch) -| (external.north);
    \draw[conn] (securityBranch) -| (internal.north);
    \draw[conn] (securityBranch) -| (empirical.north);
    \draw[conn] (securityBranch) -| (sidechannel.north);
  \end{tikzpicture}%
  }
  \caption{VR/AR software deployment}
  \label{deployment-fig}
\end{figure*}

The issue about the deployment of VR/AR software mainly consists of three parts: (1) Distribution-Platform-Related Issues; (2) Version Control and Software Updates; (3) Security and Privacy Issues. The deployment framework, together with the references that support each topic, is shown in Fig.~\ref{deployment-fig}.


\subsection{Distribution-Platform-Related Issues}
Nowadays, VR/AR applications have typically been deployed on several representative software distribution platforms, such as SteamVR, Meta Store, and PICO Store~\cite{Maxwell2022}. Since these platforms are operated by different service providers, third-party advertising partners and third-party IT-service providers are also different. Consequently, privacy policies and user data usage rules vary as the data flow generated by diverse VR/AR apps is directed to different third-party service providers across platforms~\cite{Karoline2024}. To further investigate data flows and privacy policies, \cite{Trimananda2022} designed OVRSeen with a focus on Oculus VR apps. They provided the first comprehensive analysis of personal data exposed by OVR applications and the platform itself, combining perspectives on network traffic and privacy policies. The study found that OVR is more centralized than mobile and other app ecosystems and is driven primarily by tracking and analytics rather than third-party advertising. The types of data exposed by VR apps include personally identifiable information (PII), device information that can be used for fingerprinting, and VR-specific data types. Meanwhile, \cite{zhan2024} described a system called VPVet, which was used to automate the review of VR applications for privacy policy compliance issues. It first analyzed the availability and completeness of VR privacy policies and then refined its analysis based on three key criteria: granularity, minimization, and consistency of collection/use/sharing statements. The system targeted apps on selected VR platforms, such as Meta Quest, Sidequest, and App Lab. Extending this line of research to a broader ecosystem, Meng et al.~\cite{Meng2026} conducted a large-scale longitudinal analysis of privacy policies across 10 mainstream VR distribution platforms, evaluating over 11,000 VR apps. Their empirical findings revealed severe, long-standing privacy issues in deployed apps, including inadequate policy availability, coarse-grained data-collection statements, and significant discrepancies between stated privacy policies and actual code behaviors, emphasizing the urgent need for stricter platform-level deployment regulations.
Furthermore, \cite{Yang2023} studied the security of permission management and user privacy on Android AR apps.

\subsection{Version Control \& Software Updates}
After VR/AR apps have been released to specific platforms, version control and software updates become an important issue. Distinct from traditional software, this phase introduces unique challenges, which can be categorized into the following critical research directions:

\begin{itemize}[leftmargin=*, labelindent=0pt,noitemsep,topsep=0pt]
\item \textit{Compatibility and Firmware Evolution}: Ensuring compatibility across new versions of VR software, diverse hardware, and underlying operating systems is a major challenge. Compatibility issues arising from version updates are a widely publicized problem. For example, developers have reported degraded performance when maintaining apps compatible with new Unity framework versions~\cite{Unity2023}, and noted that certain VR software updates break compatibility with current OS versions~\cite{Meta2023}. To systematically investigate these OS and firmware update issues, Simhadri et al.~\cite{Simhadri2025} recently performed a comprehensive longitudinal security analysis of over 300 VR firmware versions from major vendors (e.g., Meta, Pico). Their findings revealed that during major Android OS upgrades, deployed VR firmware frequently suffers from delayed adoption of upstream Linux kernels and inconsistent binary hardening, highlighting severe vulnerability windows during system updates.

\item \textit{Automated Patching and Maintenance}: As versions iterate, manually locating and fixing bugs becomes increasingly costly. To streamline the update process and rapidly deploy patches, Wu et al.~\cite{wu2026xrfix} introduced XRFix, an automated program repair (APR) framework powered by LLMs. This tool assists developers in automatically generating reliable patches for performance bugs in deployed applications, significantly enhancing the efficiency of version maintenance.

\item \textit{Incremental Updates}: A key open question is how a version control system should handle the massive number of binary files (e.g., 3D models, high-resolution textures, and spatial audio) and how to design efficient incremental updates to minimize downloading time and storage overhead.

\item \textit{User Experience, Data Management, and Documentation}: Several other critical issues remain actively discussed. These include Influence to User (keeping user experience consistent and avoiding discomfort during upgrades), Data Management (handling the migration, security, and integrity of user biometric data between versions), and Document Updates (providing clear change logs and user education for new VR interaction features).
\end{itemize}

Although version control and software updates for VR/AR were historically limited to community discussions, they are rapidly becoming a highly promising and rigorous research direction, as indicated by the emergence of LLM-driven automated patching and longitudinal firmware auditing.

\subsection{Security \& Privacy Issues}

In the phase of VR/AR software deployment, most research focuses on the security and privacy issues of VR/AR projects. Specifically, they mainly consist of security and privacy issues caused by external program implanting, program internal design, side-channel attacks, and issue-specific empirical studies. To systematically understand the complex threat landscape of deployed VR/AR software, Teymourian et al.~\cite{teymourian2025sok} proposed a comprehensive ontology and analysis framework for MR deception attacks, systematically mapping how attackers manipulate information channels and human cognition. To proactively mitigate these threats before final deployment, Dennis and Enes~\cite{Dennis2025} developed an immersive VR environment utilizing the ``code city'' metaphor. This tool enables developers to collaboratively visualize and explore static analysis security testing (SAST) results in 3D, facilitating thorough security audits before release.

\subsubsection{Issues Caused by External Program Implantation} \label{external}
Several recent studies have focused on attackers planting malicious code or modules in VR/AR projects to cause security threats, such as privacy breaches. Lee et al.~\cite{Lee2021} pointed out that VR advertising (ad) has become a new advertising channel with the development of WebVR technology. However, the lack of effective mechanisms to isolate the execution of ad scripts in the WebVR environment may lead to ad fraud by malicious ad servers by implanting malicious components. They introduced the following four new ad fraud attack methods: Gaze Cursor-jacking Attack, Controller Cursor-jacking Attack, Blind Spot Tracking Attack, and Abuse of an Auxiliary Display Attack. Moreover, Zhang et al.~\cite{Zhang2023} demonstrated the vulnerability of AR/VR systems to \textit{malware-initiated} side-channel attacks, where malicious applications can infer private user information without special permissions. The researchers developed a side-channel attack \textit{software} for different types of private information, including successful recovery of victims' hand gestures, voice commands, and keystrokes input on virtual keyboards. Further, \cite{yang2024} described an attack called the inception attack, in which attackers control and manipulate the user's interaction with their VR environments and applications, trapping the user in a malicious VR application that masqueraded as a complete VR system. The researchers implemented this attack and tested it on all Meta Quest VR headsets. In addition, \cite{Lehman2022} focuses specifically on the privacy threat of ``hidden operations'', where malicious or incompetent developers may perform additional visual operations modules in the background of an AR app without the user's knowledge. They proposed a framework for application development and runtime permissions that target preventing the execution of hidden operations. Through experimental results, the feasibility and usefulness of this system for distinguishing between AR applications that meet user expectations and those that do not in runtime testing were explored. 

Besides the above studies, a few studies have investigated issues with AR software. Cheng et al.~\cite{Cheng2024} explored user interface (UI) security for AR platforms. The researchers identified three attributes related to UI security: Same Space, Invisibility, and Synthetic Input, and demonstrated the impact of different implementations of these attributes on security through five proof-of-concept attacks, including clickjacking attacks and object erasure attacks. They empirically investigated five current AR platforms (ARCore, ARKit, HoloLens, Oculus, and WebXR) and found that all platforms make at least three of our proof-of-concept attacks be successful. Potential future defenses were also discussed. The work~\cite{Mengascini2024} introduced the concept of the metaverse and how it can be implemented through JavaScript programs in modern web browsers. They present a threat model that considers malicious users who exploit vulnerabilities in the metaverse platform to execute attacks by planting malicious JavaScript code. These attacks include, but are not limited to, clickjacking, user input denial of service, input forgery, object man-in-the-middle attacks, and object erasure attacks. Furthermore, the deployment of WebVR applications is highly susceptible to malicious third-party implantations, particularly within the advertising ecosystem. Mukherjee et al.~\cite{mukherjee2025shadowed} systematically investigated UI attacks in WebXR, revealing how malicious advertisers exploit VR UI properties (e.g., absence of iframe isolation, transparent meshes, sequential rendering) to inject ``dark patterns''. These external implantations covertly manipulate users into unintentional interactions, leading to ad fraud and UI-based privacy leakage without disrupting the primary app functionality.

\subsubsection{Issues Caused by Program Internal Design}
Several studies have focused on security and privacy risks in the internal design and use of VR software. Carter et al.~\cite{Carter2024} investigated shared-state attacks in multi-user AR applications based on a unified understanding of the virtual environment and its augmented ``shared state''. Current methods for creating and accessing these shared states involve collecting sensor data (e.g., camera images) from devices, processing this data, and integrating it into the shared state. However, \textit{the design of this sharing process introduced new vulnerabilities and opportunities for attacks.} This research presented a series of innovative attacks targeting various AR frameworks that utilize shared states. The study revealed that, despite the differences in their underlying mechanisms for reading and writing shared states, these frameworks all exhibited a common vulnerability to the unified threat model. Meanwhile, the work \cite{Nair2023} explored the privacy risks posed by adversarial design in VR games. By having 50 participants play a seemingly innocuous ``escape room” VR game, the researchers showed how an adversarial program can accurately infer more than 25 personal data attributes in a matter of minutes, including height, wingspan, age, gender, and more. The paper highlighted privacy risks in VR software design and encouraged privacy practitioners to focus on the intersection of data privacy and VR-enhanced social networking. 

Moreover, \cite{Samaikya2022} proposed a new framework using Attack-Fault Trees (AFTs) to quantify and analyze security and privacy issues raised through immersive attacks (e.g., occlusion attacks, overwriting attacks) and other types of attacks/faults (e.g., unauthorized access, network congestion) and demonstrated how these attacks can trigger users' motion sickness symptoms by using a social VR system. It proposed a new analytical approach to security and privacy issues in the design of social VR environments. The work~\cite{Kim2023} pointed out that existing AR platforms are not fine-grained enough in the design of access control mechanisms to effectively restrict application access to sensor data. Therefore, a framework called Erebus was proposed. Erebus enabled fine-grained control of AR application data access by creating a Domain Specific Language (DSL). The framework allowed applications to specify the data required for their functionality and allowed users to customize application permissions based on specific conditions. Further, \cite{Ruth2019} analyzed the security risks of multi-user sharing design. Zuo et al.~\cite{Zuo2022} identified the security risks of payment module design. 
In addition to data access control, establishing secure internal authentication mechanisms is vital for deployed applications to prevent identity theft. Traditional password inputs are cumbersome and vulnerable to observation in VR. To address this internal design flaw, Pan et al.~\cite{Pan2025} introduced PipID, a lightweight authentication approach leveraging built-in VR eye trackers. By analyzing the user's unique pupillary responses to randomized light stimuli on the screen, this design achieves secure, user-effortless authentication that naturally resists credential replay attacks.

\subsubsection{Issue-specific Empirical studies}
Recent empirical studies have systematically evaluated the security and privacy landscape of deployed VR applications from multiple perspectives, including app-level assessments, post-incident forensics, and human-centric developer studies.

\begin{itemize}[leftmargin=*, labelindent=0pt,noitemsep,topsep=0pt]
\item \textit{App-level Security Assessments}: Early evaluations primarily focused on broad security reviews. For instance, Noah et al.~\cite{noah2022} assessed popular AR/VR technologies by focusing on device authentication, user profiling, and access control, while \cite{Pınar2022} specifically compared the security of the main authentication methods in VR environments. To systematically scale up vulnerability evaluations, Guo et al.\cite{Guo2024} proposed a preliminary assessment tool integrating static analysis and privacy-policy analysis on 500 Oculus VR apps. Recently, they significantly extended this work~\cite{Guo2025} by proposing the full-fledged VR-SP detector and conducting a massive empirical study on 900 popular Meta VR apps. By incorporating third-party library (TPL) analysis and user review mining, they systematically exposed prevalent deployment issues, including missing root detection and unauthorized biometric data collection that severely contradicts stated privacy policies.

\item \textit{Human Factors in Deployment}: Beyond code and policies, recognizing that human factors heavily influence secure deployments is crucial. Cai et al.~\cite{Cai2020} conducted a developer-centered qualitative study with professional XR developers. They revealed a critical ``awareness gap'' where developers often underestimate XR-specific threats, and indicated that existing mitigation strategies frequently sacrifice application utility, thereby underscoring the urgent need for developer-friendly security guidelines.

\item \textit{Dynamic Exploration and Forensics}: To monitor deployed apps dynamically in the wild, Kim et al~\cite{kim2025autovr} proposed AutoVR, an automated UI exploration framework that bypasses traditional 2D limitations to dynamically execute generative events in Unity VR apps, successfully exposing hidden sensitive data flows. Furthermore, in the event of a security breach, post-incident forensics is critical. While early preliminary studies, such as Yarramreddy et al.~\cite{Ananya2018}, explored forensic analysis of VR social applications to reconstruct data from network traffic, modern complex attacks demand robust system-level auditing. To this end, Shoaib et al.~\cite{shoaib2025principled} developed a provenance-based auditing system tailored for AR/VR ecosystems. By applying NLP to unstructured logs and employing VR-aware execution partitioning, their tool allows investigators to accurately trace the root causes of complex attacks (e.g., spatial boundary manipulation) while significantly reducing graph dependency explosion.

\end{itemize}

\subsubsection{Issues Caused by Side-channel Attack}
Recently, several studies have investigated security threats and privacy risks caused by side-channel attacks. For example, \cite{Meteriz2022} discussed a keystroke logging inference attack against air-tapping keyboards in virtual environments. The researchers observed that the user's hands follow a specific pattern when typing in the air and utilized this observation to execute the attack. Notably, this paper presents an attack flow, through which the attacker can reconstruct the user's input. Meanwhile, \cite{Luo2022} described a system called HoloLogger, which can infer user inputs by monitoring the head movements of an MR HMD. They proposed a target key tracking scheme based on 6DoF head motion and a keystroke inference framework based on aerial tapping patterns. Moreover, \cite{Nahuel2021} explored an inverse problem based on keystroke timing dynamics, i.e., guessing the input text when only keystroke timing is available. The researcher proposed a method to determine whether a medium-sized list of candidate texts contains text that matches the time sequence by observing the keystroke time sequence without the need for a target text, or when the list of candidate texts is short. Further, \cite{Arafat2021} described a virtual keystroke recognition method called VR-Spy, which utilized channel state information (CSI) of WiFi signals to recognize virtual keystroke operations in VR headsets. The researchers proposed a novel pattern extraction algorithm that used signal processing techniques to extract patterns from changes in CSI and implemented an end-to-end virtual keystroke recognition model.

Besides typing inference, other studies have investigated other types of side-channel attacks. In particular, \cite{Falk2021} examined a VR de-anonymization attack technique called ReAvatar, which can identify users by analyzing their specific movement patterns in the virtual environment. Using 3D pose estimation and machine learning models, the researchers were able to relate the user's virtual character movements to real-world movements to infer the user's identity. The research also discussed how this attack could be defended against, for example, by dynamically positioning the keyboard. To safeguard the VR/AR systems from side-channel attacks, several research attempts have been made. The work~\cite{luo2020} presented a system called OcuLock that utilizes the human visual system (HVS) for authentication of VR HMDS. Unlike traditional eye-movement-based authentication methods, OcuLock considers the entire HVS, including the eyelids, extraocular muscles, cells, and peripheral nerves.  OcuLock achieves authentication based on an electro-ocular graph (EOG) HVS sensing framework with record comparison.

\summary{\revise{Software deployment is a critical phase encompassing platform distribution, version control, and security. Recent advancements have shifted from early community discussions to rigorous, automated paradigms. For version control and updates, researchers are now leveraging longitudinal firmware auditing and LLM-driven automated patching to address OS compatibility and maintenance costs. In terms of security and privacy, the research focus has significantly expanded from localized UI and side-channel attacks to large-scale empirical studies (e.g., massive NLP-based privacy policy vetting on thousands of apps), developer-centered human factor analyses, and dynamic post-incident forensics. Together, these advancements bridge the gap between abstract security guidelines and practical, reliable deployment in the fragmented VR/AR ecosystem.}}

\section{Maintenance \& Updates}
The maintenance and update phase is an ongoing post-deployment process of VR/AR software, during which developers not only respond to users' community feedback to continuously improve software quality, but also adapt to rapidly evolving technologies (e.g., generative AI) and deliver fresh content to sustain user engagement. Existing studies mainly address these challenges from two perspectives: feedback-driven continuous improvement and refactoring, and rapid evolving adaptation for long-term engagement.

\subsection{Continuous Improvement \& Refactoring via Feedback} \label{sec:Continuous Improvement}

To maintain user engagement and prevent app abandonment, developers must continuously address performance issues that can severely degrade immersive user experience, especially in VR, where problems such as cybersickness are common.  In practice, identifying and resolving these issues heavily relies on user feedback and community knowledge.

\begin{itemize}[leftmargin=*, labelindent=0pt,noitemsep,topsep=0pt]
\item \textit{Performance Optimization from Community Feedback}: During software maintenance, developers frequently rely on bug reports and discussion forums to locate performance bottlenecks. By analyzing user feedback and developer discussions (e.g., UE AnswerHub), Hogan et al.~\cite{hogan2022} showed that maintaining Unreal Engine XR projects often requires both configuration-level tuning (e.g., display settings) and C++-level  optimizations (e.g., reducing motion controller latency). Similarly, by analyzing fixing commits from open-source repositories, Nusrat et al.~\cite{9402052,Nusrat2020} found that VR performance-related maintenance in Unity commonly involves graphics simplification, rendering optimization, and heap avoidance. Notably, addressing this feedback is uniquely complex because bug fixes often require modifying not only source code but also underlying 3D assets and corresponding prefab files.

\item \textit{Detecting and Refactoring Bad Smells}: Beyond issue fixing,  maintenance also requires systematically addressing accumulated technical debt. 
To this end, Borrelli et al.~\cite{Antonio2020} introduced UnityLinter, a tool for automatically detecting Unity-specific bad smells that affect maintainability, performance, and behavior. Their survey feedback from 68 developers revealed a unique trade-off in VR software maintenance: while performance-related smells (e.g., heavyweight \texttt{\small Update()} methods) are strictly refactored to maintain a smooth user experience, maintainability smells (e.g., coupling objects through the IDE inspector) are often tolerated for editing convenience.
\end{itemize}

\subsection{Adapting to New Technologies \& Maintaining Engagement}

Besides bug fixing and refactoring, maintaining long-term user engagement and interest in VR applications also requires continuously delivering novel and appealing content. To handle the high cost of manual 3D modeling, updating VR scenes has rapidly adopted new technologies (e.g., generative AI).

\begin{itemize}[leftmargin=*, labelindent=0pt,noitemsep,topsep=0pt]

\item \textit{AI-Assisted Immersive Scene Updating}: Updating complex 3D environments (e.g., NeRFs and 3D Gaussian Splatting) to keep users engaged traditionally demands high expertise. To seamlessly adapt to generative AI capabilities, Vachha et al.~\cite{Cyrus2025} proposed Dreamcrafter, an immersive VR editor that unifies direct manipulation with generative AI by allowing developers to iteratively update and stylize 3D Radiance Fields using natural language. To preserve workflow fluidity despite the latency of AI models, Dreamcrafter introduces fast 2D proxy representations to preview updates, significantly lowering the barrier to keep the visual simulation of virtual worlds.

\item \textit{Co-Creative AI Representation}: When adapting AI as a co-creator for iterative VR content updates and sustaining user interest, how the AI interacts with users becomes a crucial task. Rasch et al.~\cite{Rasch2025} investigated different AI representation modes during continuous 3D object co-creation in VR. They found that using an embodied AI avatar to perform updates enhances the perceived partnership and support, making the updating process itself engaging. However, developers must balance this with the user's sense of ownership. Furthermore, for efficient content maintenance, the immediate visualization of AI-generated modifications is preferred over incremental visualization.
\end{itemize}

\summary{\revise{Software maintenance and updates in VR/AR are critical for preserving software quality and sustaining long-term user engagement. Recent research has evolved from ad hoc feedback handling toward more systematic, tool-supported, and AI-augmented maintenance paradigms. For continuous improvement and refactoring, researchers are leveraging community feedback mining and static analysis tools to identify VR-specific performance issues and bad smells, often requiring coordinated modifications across both source code and 3D assets. For content evolution and engagement maintenance, the research focus is increasingly shifting toward generative AI-based workflows, including immersive AI-assisted editing and co-creative AI representations that support efficient, interactive updates of high-fidelity virtual environments. Together, these advancements bridge the gap between traditional software maintenance practices and the unique demands of evolving, immersive VR/AR experiences.}}


\section{Evaluation \& Feedback}

After the deployment and maintenance phases of VR/AR software, it is vital to evaluate its quality and user experience systematically. This phase utilizes both quantitative performance measurements and qualitative user feedback to inform future iterations. The research in this phase addresses three core questions: 1) How to evaluate the app's performance in the wild? 2) How to conduct user experience and satisfaction studies? And 3) how to translate these evaluation results into actionable guidance for future development?

\subsection{Performance Measurement \& Evaluation}
Evaluating the runtime execution of deployed VR/AR applications provides objective data on the success of an application. Gathering these performance metrics is a critical first step in the evaluation phase.
\begin{itemize}[leftmargin=*, labelindent=0pt,noitemsep,topsep=0pt]

\item \textit{Evaluating Web-based XR in the Wild}: To answer how well modern frameworks support mobile XR applications, Bi et al.~\cite{Bi2023} conducted a comprehensive empirical evaluation of seven popular web-based XR frameworks (e.g., \texttt{\small Three.js}, \texttt{\small Babylon.js}) across different 3D models and platforms. By measuring First Loading Time (FLT) and Frame Per Second (FPS), their evaluation revealed that current performance bottlenecks often stem from inefficient GPU utilization and a lack of computing parallelism. These evaluation results translate into direct guidance for developers, advising them to balance 3D model complexity with real-world understanding algorithms to ensure optimal performance.
\item \textit{Evaluating via Bug Reports}: Collecting feedback from application crashes and visual glitches is another critical evaluation metric. As previously introduced in \S~\ref{sec:UsabilityTesting}, Li et al.~\cite{li2024less} proposed the StereoID framework to detect SVI. During their evaluation phase, they demonstrated the necessity of collecting and analyzing real-world bug reports across 15 different VR platforms (e.g., SteamVR and Oculus). By evaluating these reports, they successfully categorized 15 manifestations of SVI issues, directly guiding the design of their automated detection framework.

\end{itemize}

\subsection{User Satisfaction \& Experience Studies}

In addition to objective system metrics, capturing subjective user satisfaction and developer feedback is the most direct way to assess the true success of a VR/AR application and its development tools.

\begin{itemize}[leftmargin=*, labelindent=0pt,noitemsep,topsep=0pt]

\item \textit{Mining User Reviews for Cybersickness Assessment}: Manual playtesting is unscalable for large user bases and exposes testers to discomfort. To conduct large-scale user satisfaction and experience studies, Li et al.~\cite{Li2025} proposed XRCare, which is an LLM-driven framework to conduct fine-grained analysis of user reviews from VR app stores. By simulating diverse user demographics through a multi-agent system and maintaining a self-evolving reasoning graph, XRCare automatically assesses the cybersickness levels of thousands of XR apps. This innovative approach successfully translates massive, unstructured user feedback into structured causes of sickness, providing developers with actionable insights to refine user comfort.
\item \textit{Developer Surveys for Validation}: In addition to end-users, gathering feedback from developers is essential for evaluating software engineering tools. As discussed in \S~\ref{sec:Continuous Improvement}, Borrelli et al.~\cite{Antonio2020} validated their bad smell detection tool (namely UnityLinter) by conducting a user survey with 68 professional video game developers. This feedback study evaluated the perceived relevance of different bad smells, revealing that developers prioritize resolving performance smells over maintainability smells. Such developer-centric feedback translates directly into guidance for designing more practical VR linting tools.
\end{itemize}

\summary{\revise{ Software evaluation and feedback bridge the gap between current deployment and future iterations. Researchers evaluate VR/AR software success through comprehensive performance measurements (e.g., empirically benchmarking WebXR frameworks on mobile devices) and extensive bug report analysis. Furthermore, to conduct scalable user satisfaction studies, automated analysis of app store reviews using LLMs enables accurate cybersickness assessments without manual playtesting. Combined with targeted developer surveys, these evaluation methods effectively translate quantitative metrics and qualitative user feedback into actionable guidance for refining immersive applications.}}

\section{Conclusion and Future Directions}

This paper has presented a systematic and comprehensive survey of the software engineering practices applied throughout the VR/AR software life cycle. By structuring our analysis around the seven key stages—from requirement engineering to evaluation and feedback—we have mapped the current state-of-the-art, summarized key technical contributions, and highlighted the unique challenges that distinguish VR/AR software from traditional applications. Our review of over 110 studies reveals a field that is rapidly maturing, moving from adapting conventional 2D software engineering paradigms to developing bespoke, immersive-native methodologies.

Our synthesis of the existing literature illuminates several overarching themes. First, the unique nature of immersive interaction is a driving force across the entire life cycle. The need to manage 3D space, ensure user comfort (e.g., mitigating cybersickness), and support novel input modalities (e.g., 6DoF controller signals and eye-tracking inputs) fundamentally reshapes every stage, from UI/UX design and interaction prototyping to the emergence of new bug categories (e.g., stereoscopic visual inconsistencies) and security vulnerabilities (e.g., side-channel attacks via motion tracking).

Second, we observe a clear evolution toward automation and intelligence. Early development and testing relied heavily on manual processes and community feedback. However, recent advancements leverage static analysis, automated scene exploration, and, most notably, LLMs and generative AI. These technologies are revolutionizing the field by enabling VR-native IDEs, in-situ content generation, automated bug repair, and large-scale user experience analysis, thereby accelerating development and enhancing maintenance efficiency.

Third, the fragmentation of the VR/AR ecosystem remains a persistent challenge. The diversity of hardware platforms, SDKs (e.g., Unity and Unreal Engine), and distribution channels (e.g., Meta Store and SteamVR) complicates development, testing, and deployment, particularly concerning performance consistency, security policy enforcement, and long-term maintenance.

Based on our comprehensive review, we identify several promising directions for future research that are critical for advancing the field of software engineering for VR/AR:

\begin{itemize}[leftmargin=*, labelindent=0pt,noitemsep,topsep=0pt]
\item \textbf{Standardization and Cross-Platform Frameworks}: The current ecosystem is highly fragmented. Future work should focus on developing standardized APIs, interaction metaphors, and testing protocols that abstract away hardware- and platform-specific details. This would reduce development overhead, improve application portability, and create a more unified user experience. Research into universal rendering pipelines and adaptive UI frameworks that gracefully degrade across devices with varying capabilities is particularly needed.

\item \textbf{AI-Driven Immersive Development Environments}: While AI-assisted tools are emerging, the vision of a fully immersive, intelligent development environment is far from realized. Future research should explore the creation of end-to-end VR-native IDEs where developers can code, debug, and test entirely within a virtual space. This includes advancing multimodal AI agents that can interpret natural language commands, sketches, and gestures to generate and refactor complex 3D scenes and interaction logic, thus democratizing VR/AR creation for non-experts.

\item \textbf{Holistic Security and Privacy by Design}: Current security research often targets specific vulnerabilities (e.g., UI attacks, side-channel threats). A more holistic, ``security-by-design'' approach is imperative. This involves creating formal models for verifying the security and privacy properties of immersive interactions, developing privacy-preserving data-sharing mechanisms for multi-user applications, and establishing robust forensic tools for auditing incidents in the metaverse. Integrating privacy-enhancing technologies directly into the development frameworks would provide a stronger defense against the unauthorized collection of sensitive biometric and spatial data.

\item \textbf{Scalable and Human-Centric Quality Assurance}: Evaluating user experience, particularly comfort and engagement, remains a significant bottleneck. While LLM-based review analysis shows promise, future research should integrate real-time, in-application feedback mechanisms. This could involve using physiological sensors (e.g., EEG and GSR sensors) to objectively measure cybersickness and cognitive load, coupled with automated testing agents that can predict and identify UX flaws before deployment. Establishing standardized benchmarks and datasets for VR-specific bugs and performance smells would also be invaluable for the community.

\end{itemize}
In conclusion, the intersection of software engineering and VR/AR is a vibrant and rapidly evolving research area. By addressing the challenges of fragmentation, embracing AI-driven automation, and prioritizing security and user-centric quality assurance, the research community can build the robust foundations needed to unlock the full potential of the next generation of immersive software. We hope this survey serves as a valuable roadmap for researchers and practitioners, inspiring further innovation in this exciting domain.

\bibliographystyle{IEEEtran}
\bibliography{reference}

\end{document}


\title{Appendix}

\maketitle

\setcounter{table}{0}
\setcounter{figure}{0}
\setcounter{equation}{0}
\renewcommand{\thetable}{S\arabic{table}}
\renewcommand{\thefigure}{S\arabic{figure}}
\renewcommand{\theequation}{S\arabic{equation}}

\section{Detailed Search Strategy and String Formulation}~\label{sec:search detail}

To systematically retrieve relevant literature, we constructed a comprehensive Boolean search query. The query was formulated by taking the intersection (AND) of the software lifecycle domain and the immersive technology domain, while using union (OR) for synonymous terms within each domain. To minimize noise from unrelated disciplines (e.g., purely clinical psychology or optics hardware), we also introduced explicit exclusion criteria (NOT).

The general search string is defined as:
\begin{center}
\fcolorbox{black}{gray!10}{\parbox{\linewidth}{
[Software Life Cycle Keywords] AND [Immersive Technology Keywords]
}}
\end{center}
Specifically, the exact Boolean logic applied to the academic databases was:
\begin{center}
\fcolorbox{black}{gray!10}{\parbox{\linewidth}{
\{ (``Requirement Analysis'' OR ``Planning'' OR ...) OR (``Design'' OR ``UI/UX'' OR ...) OR ... \}

\textbf{\textit{AND}} 

\{``Virtual Reality'' OR ``Augmented Reality'' OR ``VR'' OR ``AR'' \}

\textbf{\textit{NOT}} 

\{``Clinical Trial'' OR "Optics" OR ``Battery'' \}
}}
\end{center}
This base query was syntactically adapted to meet the specific search field constraints (e.g., filtering by Title, Abstract, and Author Keywords) and wildcard rules (e.g., using test* to cover testing and tester) of each respective digital library, including IEEE Xplore, ACM Digital Library, and Web of Science.

\section{Breakdown of Lifecycle Keywords}~\label{sec:lifecycle keyword}

The detailed breakdown of the keywords used for each phase of the software life cycle is provided in Table~\ref{keyword}. We specifically included terms that reflect the unique characteristics of VR/AR software engineering. For instance, in the Software Design phase, we incorporated ``3D interaction'' and ``spatial design'', as these are critical, domain-specific metrics that deviate from traditional 2D software design.

\begin{table}[h!]
\centering
\caption{Keywords for Different Phases of the Software Life Cycle}
\label{keyword}
\begin{tabular}{ll}
\toprule
\textbf{Lifecycle Phase} & \textbf{Selected Keywords} \\
\midrule
\textbf{Requirement Analysis \& Planning} & requirement analysis, planning, elicitation, specification, feasibility, ... \\
\addlinespace 
\textbf{Software Design} & design, UI/UX, 3D interaction, spatial design, prototyping, storyboard, architecture, ... \\
\addlinespace
\textbf{Software Development} & development, programming, framework, compatibility, implementation, rendering, SDK, ... \\
\addlinespace
\textbf{Software Testing} & testing, test smell, bugs, bug detection, automated testing, QA, testcase, ... \\
\addlinespace
\textbf{Software Deployment} & deployment, security, privacy, distribution, version control, access control, ... \\
\addlinespace
\textbf{Software Maintenance \& Updates} & maintenance, refactoring, updates, user review, continuous integration, patching, ... \\
\addlinespace
\textbf{Evaluation \& Feedback} & evaluation, user survey, user experience, feedback, ... \\
\bottomrule
\end{tabular}
\end{table}

\section{Venue Selection \& Quality Assurance}~\label{sec:venue selection}
we ensured the high quality and academic rigor of the selected papers by strictly restricting our search space to top-tier venues. In the domain of software engineering, security \& privacy and HCI \& immersive technologies, peer-reviewed publications from premier conferences and journals inherently represent the highest standards of research quality.

Table~\ref{selected} categorizes the selected venues by their primary research domains and highlights their prestigious academic rankings (e.g., CCF A/B, CORE A/A*). By anchoring our literature collection to these highly-ranked venues, we guaranteed that all included studies possess solid methodologies, significant technical depth, and rigorous empirical evaluations.

\begin{table}[t!] \small
  \caption{Selected Conferences \& Journals and Academic Rankings}
  \label{selected}
  \resizebox{\linewidth}{!}{
  \begin{tabular}{cllc}
    \toprule
    \textbf{Domain} & \textbf{Selected Conferences} & \textbf{Selected Journals} & \textbf{Academic Ranking*} \\
    \midrule
    
    \tabincell{c}{\textbf{Software}\\ \textbf{Engineering}} & 
    \tabincell{l}{
        ACM SIGSOFT Symposium on the Foundation of Software Engineering (FSE)\\
        International Conference on Automated Software Engineering (ASE) \\
        International Conference on Software Engineering (ICSE)\\
        International Symposium on Software Testing and Analysis (ISSTA) \\
        IEEE International Conference on Program Comprehension (ICPC) \\
        International Conference on Software Analysis, Evolution, and Reengineering (SANER) \\
        International Conference on Software Maintenance and Evolution (ICSME) \\
        International Symposium on Software Reliability Engineering (ISSRE)\\
        Mining Software Repositories (MSR)
    } & 
    \tabincell{l}{
        ACM Transactions on Software Engineering and Methodology (TOSEM)\\
        IEEE Transactions on Software Engineering (TSE)\\
        Automated Software Engineering (ASE Journal) \\
        Empirical Software Engineering (ESE) \\
        IET Software (IETS) \\
        Information and Software Technology (IST)\\
        Journal of Software: Evolution and Process\\
        Journal of Systems and Software (JSS)\\
        Software Testing, Verification and Reliability (STVR)\\
        Software: Practice and Experience (SPE)
    } & 
    \tabincell{c}{CCF-A / CCF-B \\ CORE A*} \\
    \midrule
    
    \tabincell{c}{\textbf{Security}\\ \textbf{\& Privacy}} & 
    \tabincell{l}{
        ACM Conference on Computer and Communications Security (CCS)\\
        IEEE Symposium on Security and Privacy (S\&P)\\
        Usenix Security Symposium (USENIX Security) \\
        ISOC Network and Distributed System Security Symposium (NDSS) \\
        Symposium On Usable Privacy and Security (SOUPS)
    } & 
    \tabincell{l}{
        IEEE Transactions on Information Forensics and Security (TIFS)
    } & 
    \tabincell{c}{CCF-A \\ CORE A*} \\
    \midrule
    
    \tabincell{c}{\textbf{HCI \&}\\ \textbf{Immersive Tech}} & 
    \tabincell{l}{
        ACM Conference on Human Factors in Computing Systems (CHI) \\
        ACM International Conference on Ubiquitous Computing (UbiComp)
    } & 
    \tabincell{l}{
        IEEE Virtual Reality (IEEE VR)
    } & 
    \tabincell{c}{CCF-A \\ CORE A*} \\
    \bottomrule
    
    \multicolumn{4}{l}{\scriptsize \textit{*Note: Rankings are based on the latest China Computer Federation (CCF) recommended list and the CORE Conference Ranking.}}\\
  \end{tabular}}
\end{table}